%% file: main.tex
\documentclass[onecolumn]{aa} % for a paper on 1 column  
\usepackage{graphicx}
\usepackage{txfonts}
\usepackage{multirow} % for the tables
\usepackage{textcomp} % for texlangle and textrangle
\usepackage[colorlinks=true,allcolors=blue,linkcolor=red]{hyperref}
\usepackage{ulem}
\usepackage{booktabs}
\usepackage{bm}
\usepackage{multirow}

\newcommand{\oi}{O\,\textsc{\lowercase{I}}}

\newcommand{\re}[1]{{\textcolor{black}{#1}}}

\begin{document}

\title{Extended atomic data for oxygen abundance analyses\thanks{
The full tables of energy levels (Table \ref{tab:energy}) and transition data (Table \ref{tab:trdata}) are only available in electronic format the CDS via anonymous ftp to cdsarc.cds.unistra.fr (130.79.128.5)
or via https://cdsarc.cds.unistra.fr/cgi-bin/qcat?J/A+A/.}}

\author{W. Li\inst{1} \and P. J\"onsson\inst{2} \and A. M. Amarsi\inst{3} \and M. C. Li\inst{4} \and J. Grumer\inst{3}}

\institute{
$^{1}$National Astronomical Observatories, Chinese Academy of Sciences, Beijing 100012, China \email{wxli@nao.cas.cn} \\
$^{2}$Department of Materials Science and Applied Mathematics, Malm\"o University, SE-205 06 Malm\"o, Sweden\\
$^{3}$Theoretical Astrophysics, Department of Physics and Astronomy, Uppsala University, Box 516, SE-751 20 Uppsala, Sweden\\
$^{4}$School of Electronic Information and Electrical Engineering, Huizhou University, Huizhou, 516007, China \\
}

% These dates will be filled out by the publisher
\date{Received 8 December 2022/Accepted 3 April 2023. }

% Abstract of the paper
\abstract
{As the most abundant element in the universe after hydrogen and helium,
oxygen plays a key role in planetary, stellar, and galactic astrophysics.
Its abundance is especially influential on stellar structure and evolution, 
and as the dominant opacity contributor at the base of the Sun's convection zone
it is central to the discussion around the solar modelling problem.
However, abundance analyses require complete and reliable sets of atomic data.
We present extensive atomic data for \ion{O}{I}, by using the multiconfiguration Dirac–Hartree–Fock 
and relativistic configuration interaction methods.
Lifetimes and transition probabilities
for radiative electric dipole transitions are given and compared with results from previous 
calculations and available measurements.
The accuracy of the computed transition rates is evaluated by 
the differences between the transition rates in Babushkin and Coulomb gauges, as well as by a cancellation factor analysis. 
Out of the 989 computed transitions in this work, 205 are assigned to the accuracy classes AA-B, 
that is, with uncertainties less than 10\%, following the criteria defined by the National Institute of Standards and Technology Atomic Spectra Database. 
We discuss the influence of the new log($gf$) values on the solar oxygen abundance and ultimately advocate $\log\epsilon_{\mathrm{O}}=8.70\pm0.04$.}

\titlerunning{Extended atomic data for oxygen abundance analyses}
\authorrunning{W. Li et al.}
   \maketitle
%
%-------------------------------------------------------------------

%%%%%%%%%%%%%%%%%%%%%%%%%%%%%%%%%%%%%%%%%%%%%%%%%%

%%%%%%%%%%%%%%%%% BODY OF PAPER %%%%%%%%%%%%%%%%%%

\section{Introduction}\label{sec:intro}
Oxygen is the most abundant metal in the universe.
It is a key tracer of the evolution of galaxies
\citep[for example,][]{2022arXiv221004350R}, as
well as of the formation and characterisation of 
exoplanets \citep[for example,][]{2022AJ....164...87K}.
In the interiors of stars, oxygen is a major source of opacity;
for example in the Sun it is the dominant source near the base
of the convection zone \citep[for example][]{2015ApJS..220....2M}.
This makes the solar oxygen abundance critically important
for resolving the solar modelling problem, which describes
a significant discrepancy between theoretical predictions
of the solar interior structure inferred from
helioseismic inversions \re{compared to} standard solar models \citep[for example,][]{2017ApJ...835..202V,2021LRSP...18....2C}.

The abundance of oxygen in stellar atmospheres can be determined from 
analyses of stellar spectra. 
In AFGK-type stars, one of the most commonly used oxygen abundance diagnostics is the high-excitation \oi{} 777 nm triplet 
\citep[for example][]{2014A&A...568A..25N,2021MNRAS.506..150B}.
Other permitted atomic features are sometimes used as well: most commonly 
the \oi{} 615.8 nm \citep{2014MNRAS.444.3301K,2021A&A...655A..99D};
and for the Sun also the 
\oi{} 844.7 nm and 926.9 nm multiplets
\citep{2004A&A...417..751A,2021A&A...653A.141A,2008A&A...488.1031C}.
These are often complemented by low-excitation forbidden features,
usually the [\oi{}] 630.0 nm \citep{2015A&A...576A..89B,2021AJ....161....9F},
and for the Sun also the  
[\oi{}] 557.7 nm and 636.3 nm lines
\citep{2001ApJ...556L..63A,2008A&A...490..817M},
although at least in the solar spectrum these are significantly 
blended.  
Oxygen abundances can also be inferred from molecular diagnostics, in particular OH lines in the UV and infrared \citep{1998ApJ...507..805I,1999AJ....117..492B,2002ApJ...575..474M}, although such lines are typically more sensitive to the effects of stellar convection \citep[for example][]{2001A&A...372..601A,2021A&A...656A.113A}.

The accuracy of abundance determinations 
is critically dependent on reliability of the radiative transition probabilities. Moreover, if the assumption of local thermodynamic equilibrium (LTE) is to be relaxed,
as is necessary for the \oi{} 777 nm triplet and other high-excitation permitted atomic oxygen lines \citep[for example,][]{2015A&A...583A..57S,2016MNRAS.455.3735A},
a large set of reliable transition probabilities are needed to accurately determine the statistical equilibrium.

The vast majority of transition probabilities for atomic oxygen 
come from theoretical calculations. 
The transition probabilities and oscillator strengths of \oi{} presented in the Atomic Spectra Database 
of the National Institute of Standards and Technology (NIST-ASD; see \citealt{NIST_ASD}) \re{were compiled by \cite{1996atpc.book.....W}} based on the 
theoretical calculations from \cite{Hibbert_1991}, \citet{BUTLER1991}, and \cite{1992A&A...265..850B}. 
\cite{Hibbert_1991} computed the oscillator strengths for a large number of allowed transitions connecting the 
$n$ $\le$ 4 triplet and quintet states of neutral oxygen, using the CIV3 code.
\cite{BUTLER1991} performed the calculations of oscillator strengths for allowed transitions in \oi{}, in the framework of
the international Opacity Project.
Using the \sout{atomic structure} computer program SUPERSTRUCTURE, \cite{1992A&A...265..850B} calculated 
the oscillator strengths for 2$p$-3$s$ and 3$s$-3$p$ \re{spin-}allowed or spin-forbidden transitions of astrophysical interest.

There are also a number of other calculations available for \oi{}.
\re{The complete lists of published papers can be retrieved from the NIST Atomic Transition Probability Bibliographic database \citep{kramida.2023}.}
\cite{Tachiev2002A&A...385..716T} performed multi-configurational Hartree-Fock (MCHF) calculations including Breit-Pauli effects in subsequent configuration-interaction calculations and determined lifetimes and transitions rates for all {fine-structure} levels up to \re{${2p^33d}$} of the 
oxygen-like sequence (\re{elements with atomic number $Z$ = 8-20}). \cite{2002ApJS..143..231Z} calculated the atomic data, including the radiative lifetimes, transition probabilities, 
and oscillator strengths in \oi{}, by employing the weakest bound electron potential model theory.
Using the B-spline box-based R-matrix method in the Breit–Pauli formulation, \cite{Tayal_2009} calculated the oscillator strengths for allowed transitions among 
the \re{$n$} = 2–4 levels and from the \re{$n$} = 2 levels to higher excited levels up to \re{$n$} = 11 in neutral oxygen.

In this work, we present extended calculations of atomic data
for the lowest 81 states in \oi{}, using the multiconfiguration Dirac-Hartree-Fock (MCDHF) and relativistic configuration interaction (RCI) methods.
These calculations are part of an overarching project concerning the astrophysically
important CNO neutral elements, and extensive results have already been reported
earlier for \ion{C}{I} \citep{2021MNRAS.502.3780L} and \ion{N}{I} \citep{2023ApJS..265...26L}.
Electric dipole (E1) transition data (wavelengths, transition probabilities, line strengths, and weighted oscillator strengths) 
are computed along with the corresponding lifetimes of these states.
We then investigate how the differences in the calculated log($gf$) may influence the solar oxygen abundance and thereby the solar modelling problem.

\section{Theoretical method}\label{sec:theory}
\subsection{Multiconfiguration Dirac–Hartree–Fock approach}
The calculations were performed using the \textsc{Grasp2018}
package\footnote{\textsc{GRASP} is fully open-source and is available on GitHub repository at \url{https://github.com/compas/grasp} maintained by the CompAS collaboration.} \citep{Grasp2018,manual2022}, which is based on the MCDHF and RCI methods.
{Details} of the MCDHF method can be found in \cite{Grant2007}, \cite{fischer.2016}, \cite{Grasp2018}, and \cite{atoms11010007}. Here we only give a brief introduction.
%\jon{[jon comment: did you use CPC GRASP2018 or the current dev version at our github? This is important since they do actually differ a bit. It is also important to point users to the github repo. If you use github GRASP then I suggest you add a footnote here as well, saying e.g.: "GRASP is fully open-source and is hosted at \url{ https://github.com/compas/grasp}]}\textcolor{green}{I used GRASP2018. I added the GRASP2018 reference!}

In the MCDHF method, wave functions $\Psi$ for atomic states
$\gamma^{(j)}\, P JM$, $j=1,2,\ldots,N$
with angular momentum quantum numbers $JM$ and parity $P$ are expanded over {${N_\mathrm{CSFs}}$ configuration state functions (CSFs)}
\begin{equation}
    \Psi(\gamma^{(j)}\, P JM) = \sum_i^{{N_\mathrm{CSFs}}} c^{(j)}_i\, \Phi(\gamma_i\, P JM).
    \label{eq:csfs}
\end{equation}
The CSFs are $jj$-coupled many-electron functions built
from products of one-electron Dirac orbitals. As for the notation,
$J$ and $M$ are the angular quantum numbers, $P$ is parity, and $\gamma_i$
specifies the occupied subshells of the CSF with their complete angular
coupling tree information, for example orbital occupancy, coupling scheme
and other quantum numbers necessary to uniquely describe the CSFs.

The radial large and small components of the one-electron orbitals together with the expansion coefficients \{$ c^{(j)}_i $\} of the CSFs 
are obtained in a relativistic self-consistent field procedure, by solving the Dirac-Hartree-Fock radial equations and the configuration interaction eigenvalue problem 
resulting from applying the variational principle on the statistically weighted energy functional of the targeted states with terms added for preserving the
orthonormality of the one-electron orbitals. 
The angular integrations needed for the construction of the energy functional are based on the second quantization method in the coupled tensorial form \citep{1997JPhB...30.3747G,2001CoPhC.139..263G}
and account for relativistic kinematic effects. 
Once the radial components of the one-electron orbitals are determined, higher-order interactions,
such as the transverse photon interaction and quantum electrodynamic effects (vacuum polarization
and self-energy), are added to the Dirac-Coulomb Hamiltonian.
Keeping the radial components fixed, the expansion coefficients \{$ c^{(j)}_i $\} of the CSFs 
for the targeted states are obtained by solving the configuration interaction eigenvalue problem.

The transition data, for example, transition probabilities and weighted oscillator strengths,
between two states $\gamma'P'J'$ and $\gamma PJ$ are expressed in
terms of reduced matrix elements of the transition operator ${\bf T}^{(1)}$:
\begin{eqnarray}
\langle \,\Psi(\gamma PJ)\, \|  {\bf T}^{(1)} \| \,\Psi(\gamma' P'J')\, \rangle  &=&
\nonumber \\
 \sum_{j,k} c_jc'_k \; \langle \,\Phi(\gamma_j PJ)\, \|  {\bf T}^{(1)} \| \,\Phi(\gamma'_k P'J')\, \rangle,
 \label{eq:tr}
\end{eqnarray}
where $c_j$ and $c'_k$ are, respectively, the expansion coefficients of the CSFs for the lower and upper states. The summation runs over all basis states included in the two CSF expansions \eqref{eq:csfs}.

\subsection{d$T$ and CF}\label{accuracy} \label{sec:dT&CF}
In relativistic theory, there are two common representations of the E1 transition operator, the Babushkin and Coulomb gauge, which are equivalent to the length and the velocity forms in the non-relativistic limit. Just as for the latter two, assuming the wavefunctions to be exact solutions to the Dirac equation leads to identical values for the Babushkin and Coulomb transition moments \citep{1974JPhB....7.1458G}.
%\remove{For E1 transitions, there are two forms of the transition operator: Babushkin and Coulomb forms, which are equivalent to the length and the velocity form in non-relativistic limit. Babushkin and Coulomb gauges for the exact solutions of the Dirac-equation give the same value of the transition moment.} %\citep{1974JPhB....7.1458G}.

For approximate solutions achievable in \re{practise}, the transition moments differ and the quantity d$T$, defined as 
\begin{equation}\label{eq:dT}
\rm{d}\it{T}=\frac{|A_\mathrm{B}-A_\mathrm{C}|}{\max(A_\mathrm{B}, A_\mathrm{C})}\re{,}
\end{equation}
where $A_{\rm{B}}$ and $A_{\rm{C}}$ are transition rates in the Babushkin and Coulomb forms \citep{Fischer2009,Ekman2014}, can \re{be used} to evaluate the uncertainty of the computed rates in a statistical sense for \re{a} group of transitions.

The accuracy of the computed transition data can also be evaluated by studying the cancellation
factor (CF), which is defined as \citep{Cowan1981}
\begin{equation}\label{eq:CF}
\mathrm{CF} = \left[\frac{|\sum_j\sum_kc_j\langle \,\Phi(\gamma_j PJ)\, \|  {\bf T}^{(1)} \| \,\Phi(\gamma'_k P'J')\, \rangle c'_k|}{\sum_k\sum_j|c_j\langle \,\Phi(\gamma_j PJ)\, \|  {\bf T}^{(1)} \| \,\Phi(\gamma'_k P'J')\, \rangle c'_k|} \right]^2,
\end{equation}
where the notations are the same with those in Eqs. \eqref{eq:csfs} and \eqref{eq:tr}.
A small value of the CF, for example, less than 0.1 or 0.05 \citep{Cowan1981},
indicates that the calculated transition parameter, such as the transition rate or line strength, is affected by a strong cancellation effect.
This occurs due to configuration interaction between basis states of opposite phase but almost equal amplitudes, resulting in a relatively small line strength.
%\remove{This can happen due to the configuration-interaction effect when the contributions from the mixing configurations are with  opposite sign but almost equal amplitudes, resulting in a small oscillator strength for the transition.}
Transitions with small CFs are normally associated with 
large uncertainties; therefore the CF can be used as a complement 
to the d$T$ values for the uncertainty estimation. 
In this work, we {extend the} \textsc{Grasp2018} package
\citep{Grasp2018, manual2022}
to include the calculation of CFs.

\input{scheme}

\subsection{Computational schemes}
Calculations were performed in the extended optimal level
(EOL) scheme \citep{EOL} for the weighted average of the even states
(up to \re{${2s^22p^35f}$}) and odd states (up to \re{${2s^22p^35d}$}).
The CSF expansions were obtained using the multireference single-double (MR-SD) method, 
allowing single and double (SD) substitutions from MR configurations to orbitals in an active set (AS) \citep{olsen.1988, sturesson.2007, fischer.2016}.).
In addition to the target configurations {representing the physical states}, a number of configurations giving considerable contributions to the total wave functions are also included in the MR\re{.} 
%\jon{[comment: maybe say something about how these are determined, i.e. explain how rmixaccumulate works and what cut-off you use?]}. 
{SD substitutions from such an extended MR has the effect of including higher-order configuration-interaction contributions in the wavefunctions (relative to the target configurations).} 
The two MR sets for the even and odd parities are presented in
Table \ref{tab:MR}, which also displays the AS and the 
number of CSFs in the final even and odd state expansions
distributed over the different $J$ symmetries.

Similarly to the computational schemes used in \ion{C}{I-IV} \citep{2021MNRAS.502.3780L}
and \ion{N}{I} \citep{2023ApJS..265...26L}, following the CSF generation strategies suggested by \cite{Papoulia2019},
the MCDHF calculations were based on CSF expansions for which we
impose restrictions on the substitutions from the inner subshells
to obtain a better representation of the outer parts of the
wavefunctions of the higher Rydberg states, as a
consequence, improving the accuracy of the transition data.
In the initial calculations, we investigated the contribution of
the core-valence (CV) correlations \re{to} the results by allowing at
most one substitution from \re{${1s^2}$} and found that the CV
\re{contributions} are negligible.
Therefore, the \re{${1s^2}$} core remained frozen in both \re{the} MCDHF and RCI calculations.
The CSF expansions used in \re{the subsequent} MCDHF calculations were obtained
by allowing SD substitutions from the 2\re{$p$} orbital of the target
configurations to the active set of orbitals. During this stage,
the 1\re{s} and 2\re{$s$} orbitals were kept closed.
The final wavefunctions of the targeted states were
determined in an RCI calculation, which included CSF expansions
that were formed by allowing SD substitution from all subshells 
with \re{$n$} = 2 of the MR configurations.

\section{Results and discussions}
\subsection{Energy levels and lifetimes}

\begin{figure}
    \centering
\includegraphics[width=0.48\textwidth,clip]{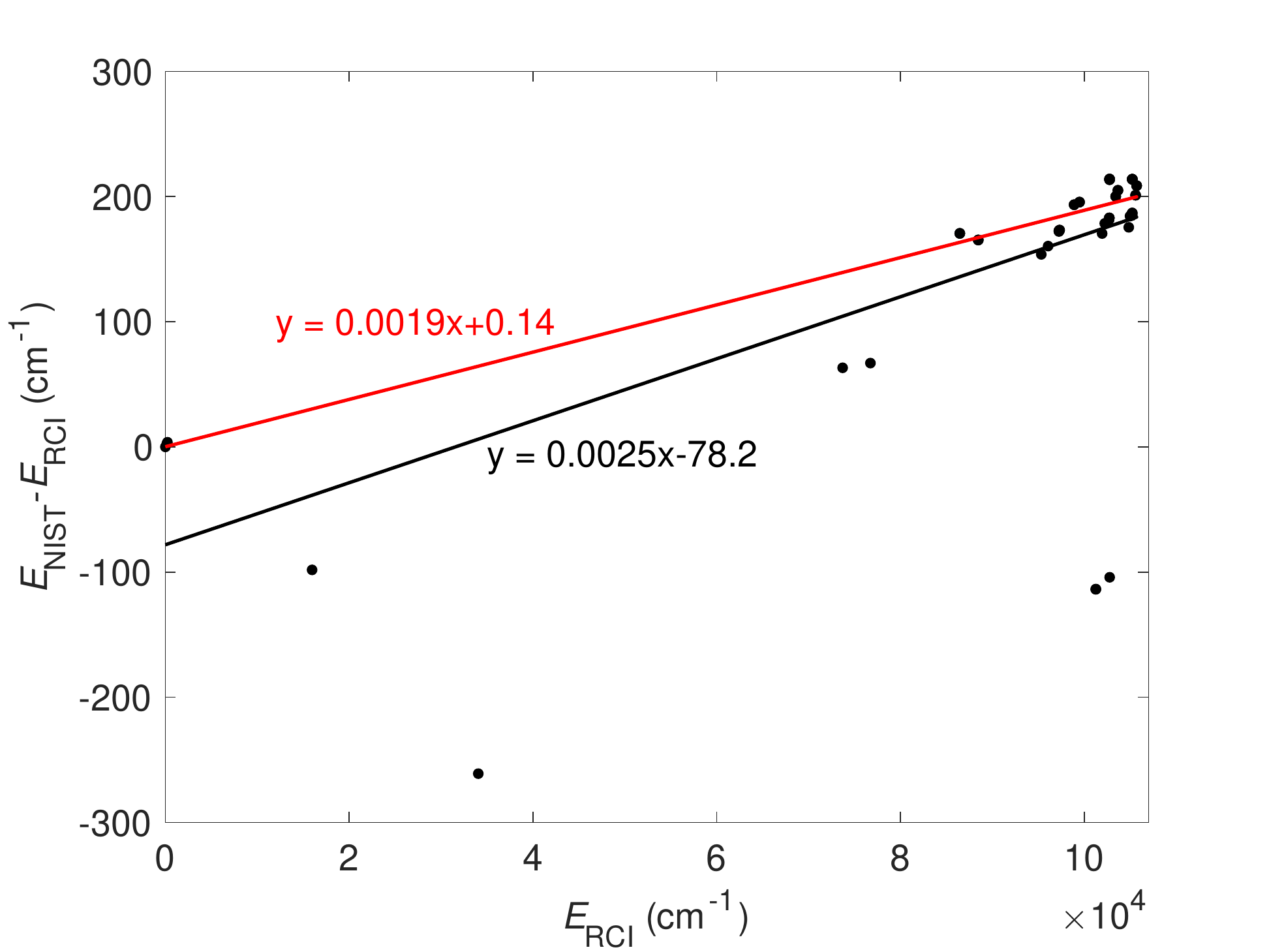}
    \caption{\re{The energy differences along with present computed excitation energies. The black solid line is the linear fit to the scatter data shown in the figure. The red solid line is the linear fit by excluding the $2s^22p^4~ ^1D_2$, $^1S_0$, $2s^22p^33s~^5S_2$, $^3S_1$, $^3D_{1,2,3}$, and $^1D_2$ states.}}
    \label{fig:energy}
\end{figure}

\re{The energies for the 81 lowest states of \oi{} (45 even states and 36 odd states)}
are given in Table \ref{tab:energy}. 
In the calculations, the labelling of the eigenstates is determined
by the CSF with the largest coefficient in the expansion of 
Eq. \eqref{eq:csfs}.
For comparison, the observed energies from the NIST-ASD \citep{NIST_ASD}, 
together with the differences $\Delta E$ = $E_{\rm{NIST}}$ - $E_{\rm{MCDHF}}$ are aslo displayed in the table.
In most cases, the relative differences between theoretical and experimental results
are less than 0.2\%, with the exception of
the levels belonging to the ground configuration \re{${2s^22p^4}$}, for which the average
relative difference is about 1.16\%.
\re{Fig. \ref{fig:energy} shows the energy differences between NIST-ASD values and the present computed data, $\Delta E$, plotted against the excitation energies, $E\mathrm{_{RCI}}$. From the linear fitting we can see that the computational excitation energies have a systematic error of 0.25\%. We observed that for most of the levels, the computed results are smaller than the NIST-ASD values by about 170-190 cm$^{-1}$ except for a few states belonging to $2s^22p^4$ and $2s^22p^33s$. By excluding these levels, that is, $2s^22p^4~ ^1D_2$, $^1S_0$, $2s^22p^33s~^5S_2$, $^3S_1$, $^3D_{1,2,3}$, and $^1D_2$, the systematic error decreased to 0.19\%.}
In the last two columns of Table \ref{tab:energy}, lifetimes obtained
from the computed E1 transition rates in both Babushkin and Coulomb gauges
are also presented. The relative differences between \re{the} Babushkin and Coulomb gauges are well below 5\%, 
except for a few states \re{for which} decay to the lower states \re{is} dominated by intercombination transitions.

In Table \ref{tab:tauexp}, the lifetimes from the present MCDHF/RCI calculations
are compared with available results from other theoretical calculations and experimental measurements.
The calculated lifetimes in \re{the} Babuskin and Coulomb forms 
are consistent to 6.0\% for all the selected transitions. 
Among others, atomic properties of \re{the} metastable state \re{${2p^33s~^5S^o_2}$} are interesting due to its 
potential in astrophysical diagnosis.
The lifetime of the \re{${2p^33s~^5S^o_2}$} state has been studied 
systematically using the MCDHF method by \cite{PhysRevA.102.042824} and \re{the final values of 
202$\pm$30 $\mu$s in the Babuskin gauge and 235$\pm$35 $\mu$s in the Coulomb gauge were} recommended.
Our calculated results of 209/213 $\mu$s (in B/C forms) are in good agreement with their
values. The much larger lifetime from the MCHF calculation by \cite{Tachiev2002A&A...385..716T} is {likely} caused by neglected
electron correlation and relativistic effects. However, the theoretical lifetimes from the various calculations
are still larger than the experimental values {reported} in \cite{1974PhRvA...9..568W}, \cite{1972PhRvA...5.2688J}, \cite{1978PhRvA..17.1921N}, and \cite{1990ch2..book.....M}, obtained using the time-of-flight technique.
It is also interesting to note that the theoretical lifetimes from \cite{Tayal_2009}, based on the B-spline box-based R-matrix method, for \re{${3s~^3D^o}$}, \re{${4s~^3S^o}$}, \re{${4d~^5D^o}$}, and 
\re{${4d~^3D^o}$} states, {differ significantly} from those obtained with the other three {theoretical approaches}, which are \cite{Tachiev2002A&A...385..716T} using the MCHF
{method implemented in the ATSP code \citep{FISCHER2000635,FROESEFISCHER2007559}, and the configuration interaction (CI) calculations of \cite{Hibbert_1991} and \cite{1992A&A...265..850B} that \re{were made with} the CIV3 \citep{HIBBERT1975141} and SUPERSTRUCTURE codes \citep{refId0}, respectively.}

There are a number of measurements of lifetimes in \oi{} presented in, for example, \cite{1977ApJ...214..328B} using the electron-beam
 phase-shift method, \cite{1985PhRvL..55..284K} from the time-resolved laser spectroscopy, and \cite{1971ApJ...165..217S,1974CaJPh..52.1961P} using the 
 beam foil technique.
The measurements of \cite{1977ApJ...214..328B} agree within the experimental errors with our calculated lifetimes for \re{3$s~{^3S^o, ^3D^o}$}, and \re{$5d~{^3D^o}$} states, while showing large discrepancies for \re{$4s, 5s~{^3S^o}$} and \re{$4d~{^3D^o}$} states.
Our lifetimes of \re{$3s~^3S^o, ^3D^o$, and $^1D^o$} states agree well with the experimental results published by \cite{1971ApJ...165..217S} using the beam-foil method; while \cite{1974CaJPh..52.1961P}, which utilized the same experimental technique, measured somewhat larger values.
Time-resolved spectroscopy and \re{high frequency deflection technique were used, respectively,} by \cite{1985PhRvL..55..284K} and \cite{1978PhyS...17..119B} for measuring the
lifetimes of \textit{ns}, \textit{np} and \textit{nd} states in \oi{}. The results from \cite{1985PhRvL..55..284K} are consistent with our theoretical lifetimes within
the experimental errors; while the results from \cite{1978PhyS...17..119B} are either too large or too small, when compared with the theoretical predicted values. 
For \re{$3s~{^3S^o}$} state, our results agree well with all the experimental data if we exclude the
lifetime from \cite{1966ApJ...146..940S}, which is evidently too large. 

\input{tau_exp}

\subsection{Transition rates and oscillator strengths}

Transition data in the form of wavelengths, line strengths (S), 
weighted oscillator strengths (log($gf$)), transition probabilities ($A$), gauge agreements, d$T$, cancellation factors, CF, as well as the estimated accuracy classes of all computed E1 transitions are given in Table \ref{tab:trdata}.
{It is important to note}
that the {reported} wavelengths {have been} adjusted to match the level energy values in the NIST-ASD. The data for log($gf$) and \textit{A} are reported in both Babuskin and Coulomb gauge and are adjusted using experimental wavelengths.

\begin{figure*}
    \centering
    \includegraphics[width=0.33\textwidth,clip]{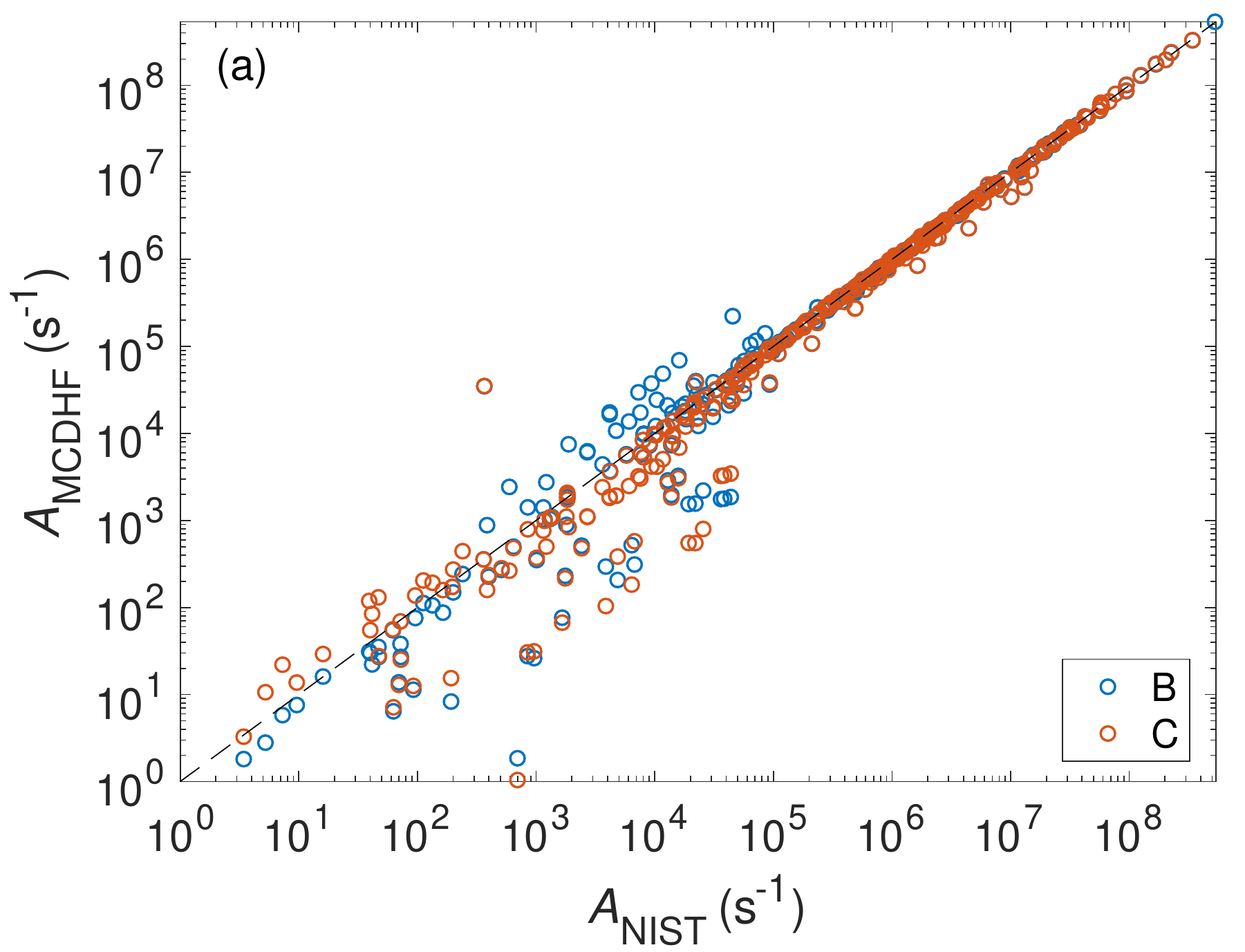}
    \includegraphics[width=0.33\textwidth,clip]{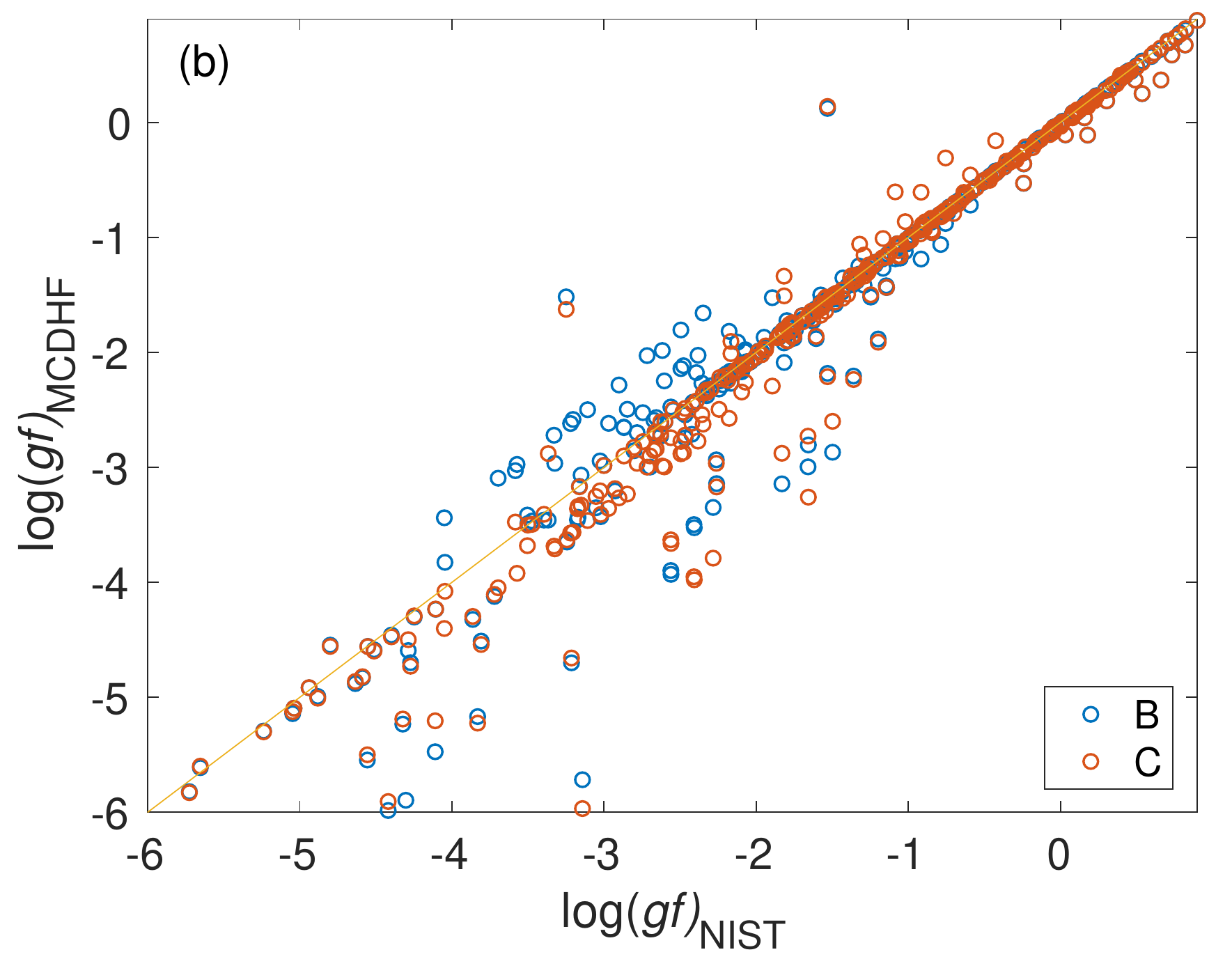}
    \includegraphics[width=0.33\textwidth,clip]{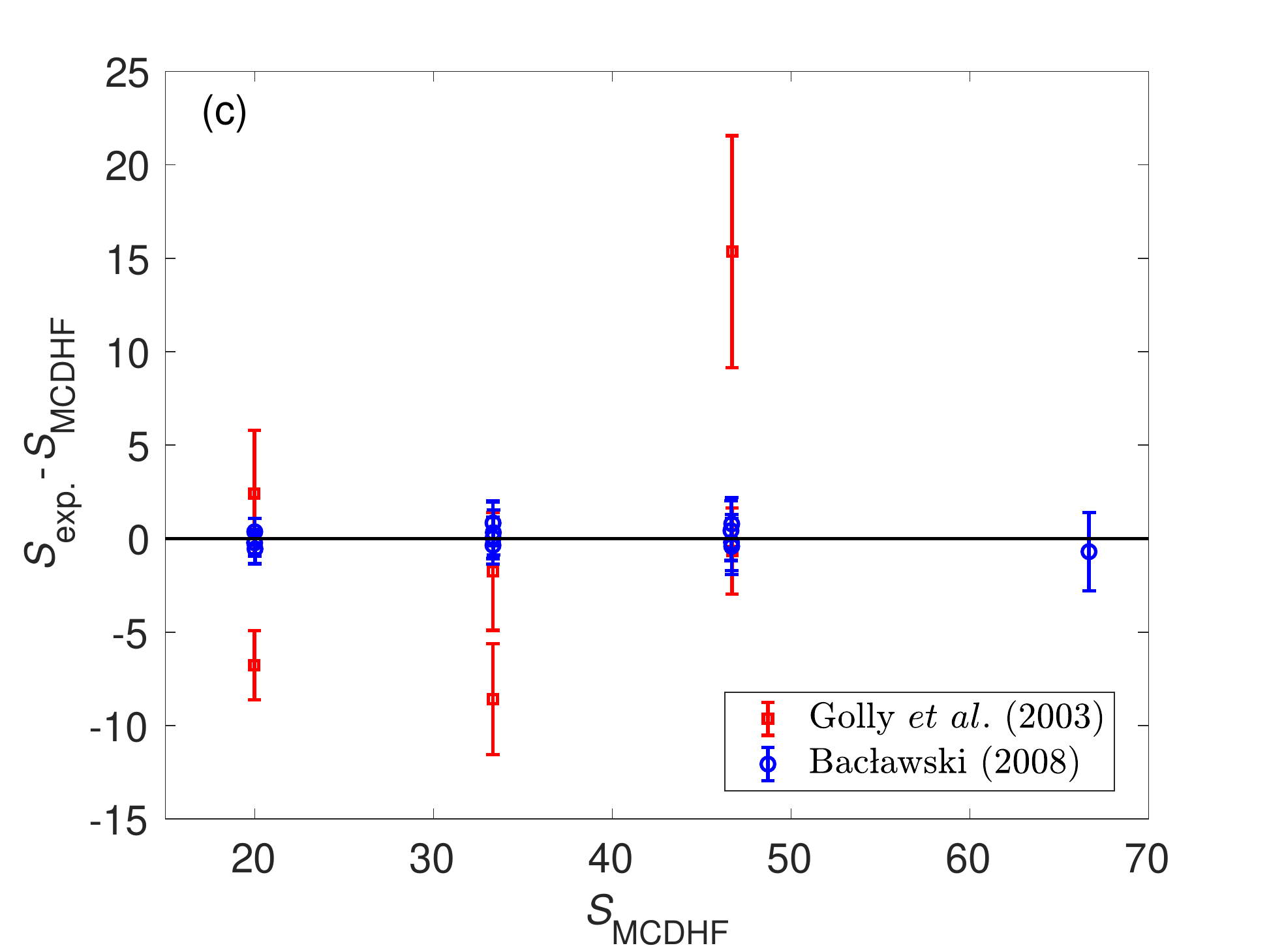}
    \caption{Comparison of transition data of the current study with the values from other work. Panel (a): Comparison of theoretical transition probabilities, $A$, in both Babushkin (B) and Coulomb (C) gauge with the results available in the NIST-ASD \citep{NIST_ASD}. Panel (b): Comparison of the log($gf$) values in Babushkin form with the results from CIV3 \citep{Hibbert_1991}, MCHF \citep{Tachiev2002A&A...385..716T} and B-spline R-matrix \citep{Tayal_2009} calculations. 
    \re{Panel (c): Comparison of the theoretical line strengths, $S\mathrm{_{MCDHF}}$, with the corresponding experimental results published by \cite{2003PhyS...67..485G} (in red square) and \cite{2008JQSRT.109.1986B} (in blue circle). Note that relative line strengths were provided by \cite{2008JQSRT.109.1986B}, which means that the line strengths are normalised within each multiplet to the sum of 100; the corresponding results from present calculations and \cite{2003PhyS...67..485G} are done with the same procedure. The line strengths in Babushkin gauge are used for the comparison.}}
    \label{fig:NIST}
\end{figure*}

\subsubsection{Comparisons with previous theoretical results}
The calculated transition results are compared with other theoretical works. 
In the left panel of Fig. \ref{fig:NIST}, $A$ values in both Babushkin and Coulomb gauges from the present work are compared with available data from the NIST-ASD \citep{NIST_ASD}, which are compiled based on the results from \cite{Hibbert_1991}, \cite{1992A&A...265..850B}, and \cite{BUTLER1991}. The {first two} calculations \re{were} carried out using the CIV3 and SUPERSTRUCTURE code, respectively. The latter work \re{was} done within the framework of the international Opacity Project.

As shown in the figure, the agreement between the $A$ values computed in the present work and the respective results from the NIST-ASD \re{is} rather good for most transitions, especially for transitions with $A_\mathrm{B}$ $\geq$ 10$^5$ s$^{-1}$; \re{taking results in Babuskin gauge as example,} 70\% of the 380 selected transitions are in agreement with the NIST-ASD data with the differences less than 20\%. \re{204 out of the 236 transitions having $A_\mathrm{B}$ $\geq$ 10$^5$ are in agreement with the NIST-ASD data with the relative differences less than 10\%.}
It is interesting to note that, for transitions with 10$^3$ $\leq A \leq $ 10$^5$ s$^{-1}$, the transition data in \re{the} Coulomb \re{gauge} are more consistent with the results in the NIST-ASD than those in \re{the} Babushkin \re{gauge}.
On closer inspection of these transitions we found that most of them are transitions involving high Rydberg states. For example, 74\% of them are those from $n$ = 5, 6 states to lower levels. For {this class} of transitions we recommend the radiative data {calculated} in \re{the} Coulomb gauge, and not the more conventional Babuskin gauge.
Following \cite{Papoulia2019}, the reason {for this} is that correlation orbitals resulting from MCDHF calculations based on CSF expansions obtained by {including} substitutions from deeper subshells are contracted in comparison with the outer Rydberg orbitals. As a consequence, the outer parts of the wavefunctions for the {relatively extended} Rydberg states are not accurately described.
%\remove{Thus, the length form that probes the outer part of the wavefunction does not produce trustworthy results, while the velocity form that probes the inner part of the wavefunctions yields more reliable transition rates.}
{Thus, it can be argued that the Coulomb gauge, which is weighted on the inner parts of the wavefunction, should yield more reliable transition data than the Babushkin gauge for transitions involving high-lying Rydberg states.}
\re{However,} for transitions involving low-lying states, our calculated transition rates in two gauges are very consistent \re{with} each other, except for some weak transitions with $A$ < 10$^2$ s$^{-1}$; for these transitions with large differences between two gauges, the Babushkin form is generally preferred\re{,} since it is more sensitive to the outer part of the wave functions that governs the atomic transitions \citep{1974JPhB....7.1458G, AHibbert_1974}.
%\jon{[comment 1: there's one step in the argument missing here? The B gauge is generally more accurate for low-lying states since it is weighted on the outer part of the wavefunction, as compared to the C gauge, which is also where our attention in the VV-correlation model is put? Or, how would you explain it?]}
%\jon{[comment 2: it would be really nice to be able to distinguish between Rydberg and low-lying transitions in Fig 1....but we can save that for another day :)]}
%\amat{any recommendation on which gauge to use for the other transitions, i.e. not the ones involving Rydberg states?} \textcolor{green}{WX:if dT is rather small, then either is fine. I would prefer length gauge for transitions involving low lying states. I added one sentence here.} \per{PER: I agree}

Furthermore, in the middle panel of Fig. \ref{fig:NIST}, the computed log($gf$) values in the Babushkin gauge are compared with the results from various calculations of \cite{Tayal_2009}, \cite{Tachiev2002A&A...385..716T}, and \cite{Hibbert_1991}, which were carried out by B-spline R-matrix, CI and MCHF methods, respectively.
From the figure, we note an excellent agreement between the present results and \re{the} other three theoretical values for transitions with log($gf$) > $-$1.5. However, for weaker transitions with log($gf$) < $-$1.5, agreement between different methods is \re{worse} with a much wider scatter.  
Overall, the log($gf$) values from the present work are in better agreement with those from \cite{Tachiev2002A&A...385..716T} and \cite{Hibbert_1991} than those from \cite{Tayal_2009}.

\input{gf_exp}
\subsubsection{Comparisons with available experimental results}
In Table \ref{tab:gfexp}, the computed $gf$ values are compared with \re{some} available results from experimental measurements.
\cite{Goldbach1994A&A...284..307G} measured the log($gf$) values of 12 lines of \oi{} belonging to 5 multiplets in the 95-120 nm %\amat{i suggest to use only \AA or nm throughout the paper... in the introduction we use nm.}\textcolor{green}{WX:Done!}
spectral range with a wall-stabilized arc. The uncertainty achieved in the measured absolute $gf$-values is between $\pm$10 and $\pm$20\%.
Our computed $gf$ results in both gauges are in excellent agreement with the measured values by \cite{Goldbach1994A&A...284..307G}, except for 
the 115.215 nm and 99.080 nm lines, for which we predict slightly larger values (by 2\%). For the 115.215 nm line, other determinations of $gf$ value have also been achieved by utilizing various techniques, for instance, 
beam-foil technique \citep{1971ApJ...165..217S,1971JOSA...61..519M,1972CaJPh..50.2496L,1974CaJPh..52.1961P}, phase-shift technique \citep{1968ApJ...152..695G}, or pulsed electron beam \citep{1970PhRvA...2..397L}.
From Table \ref{tab:gfexp} we notice that different methods obtained very consistent results, and our computed values are in nice agreement with them. 
A number of measurements of $gf$ values have also been done for the 130.6 nm line. Again, the results from different measurements agree perfectly with each other, as well as with our computed results in both Babushkin and Coulomb gauges.
\re{In addition, \cite{1998PhRvA..57.4960B} measured the transition probabilities of the $2p^33s~^3S^o - 4p~^3P$ and $3s~^5S^o - 2p^34p~^5P$ arrays and obtained the values of 7.62 $\times 10^5$ s$^{-1}$ and 3.64 $\times 10^5$ s$^{-1}$, respectively. Our computed results are in agreement with the experimental value for the $2p^33s~^3S^o - 2p^34p~^3P$ array within the experimental error, while predict slightly larger transition probability for the $2p^33s~^5S^o - 2p^34p~^5P$ array, that is, by about 5\% after considering the experimental uncertainty.
}

\re{There are also measurements of line strengths for spectral lines in visible and infrared~\citep{2003PhyS...67..485G,2008JQSRT.109.1986B}. In Fig. \ref{fig:NIST} (c), the computed $S$ values in Babushkin gauge are compared
with experimental results from \cite{2003PhyS...67..485G} and \cite{2008JQSRT.109.1986B}, by plotting line strength differences {}{versus} the computed line strengths.
Note that \cite{2008JQSRT.109.1986B} provided the relative line strengths within multiplets (normalised to 100); therefore all the values used for comparison in Fig. \ref{fig:NIST} (c) are converted to the relative values.  
We can see that all of our computed lines strengths are in agreement with the results from \cite{2008JQSRT.109.1986B} within the reported experimental uncertainties. However, in comparison with the results by \cite{2003PhyS...67..485G}, large discrepancies are observed for the $2p^33p~^5P - 2p^33d~^5D^o$ transition array, for which our computed results are in perfect agreement with those from \cite{2008JQSRT.109.1986B}. }

\subsubsection{Uncertainty estimation using d$T$ and CF}\label{uncertainty}

\input{statistic}

\input{accuracy}

\begin{figure*}
    \centering
    \includegraphics[width=0.48\textwidth,clip]{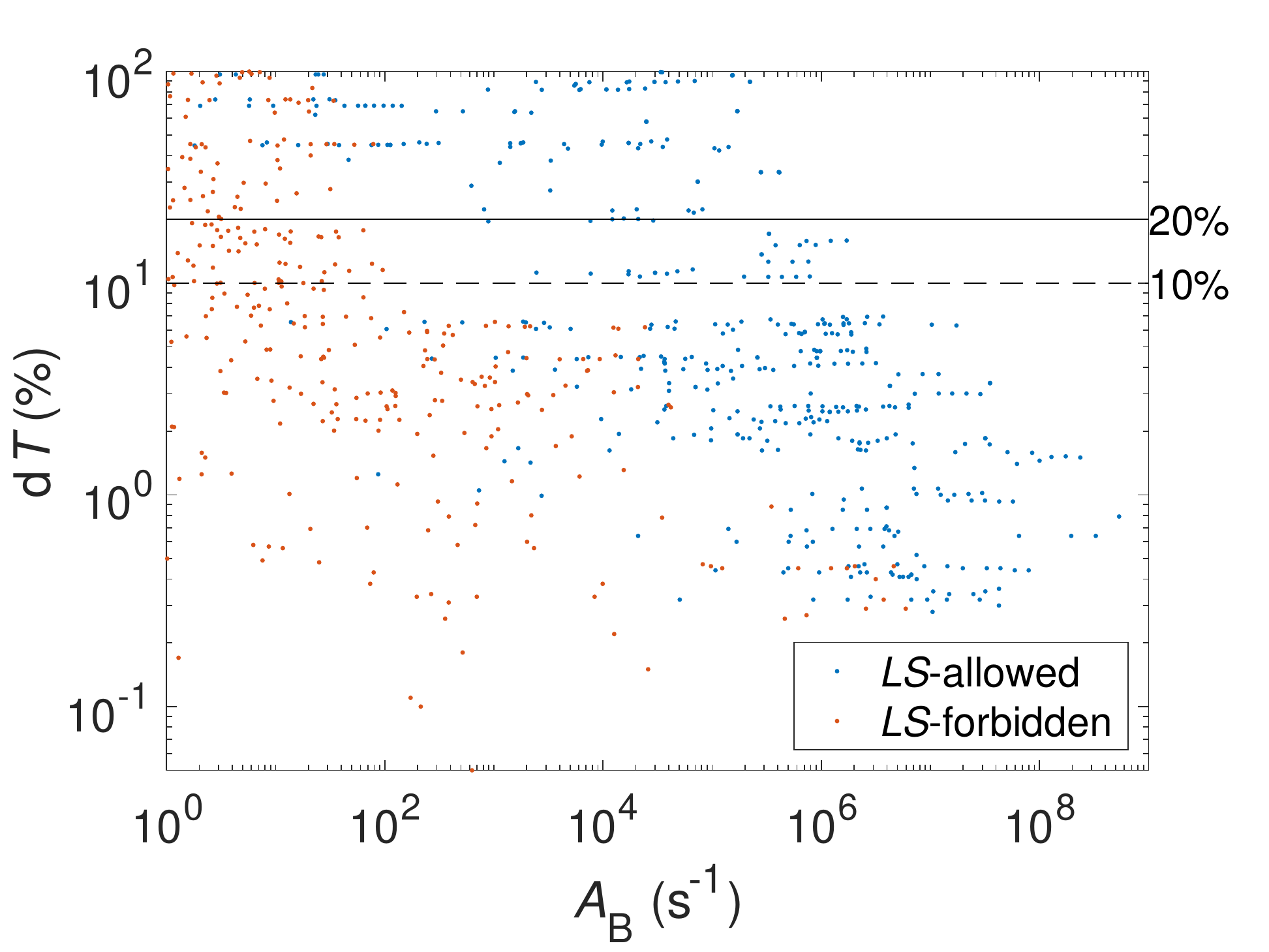}
    \includegraphics[width=0.48\textwidth,clip]{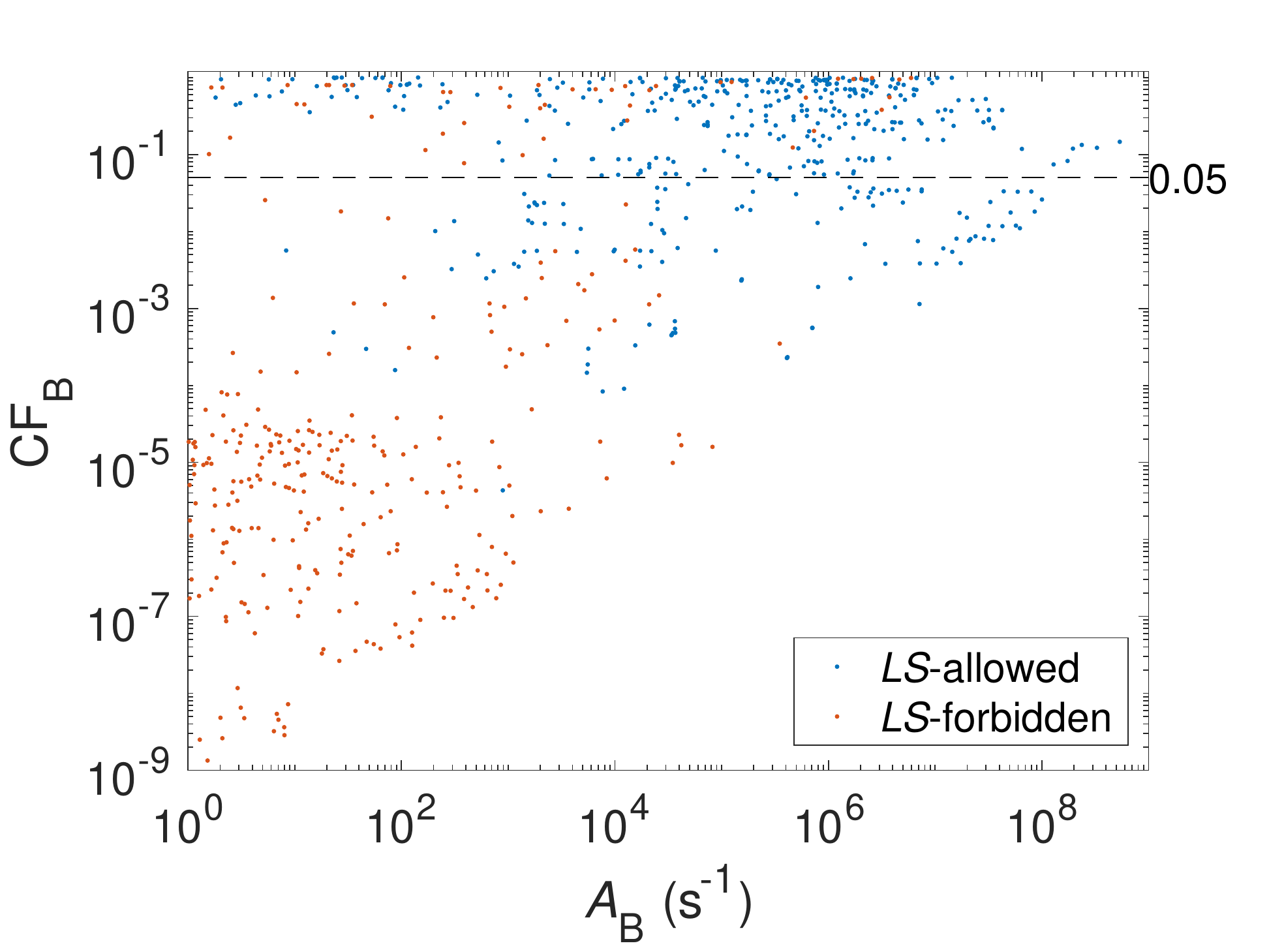}
    \caption{Scatterplot of d$T$ and CF values versus transition rates $A$ of E1 transitions, for \oi{}. Left panel: Scatterplot of d$T$ values versus transition rates in Babushkin form, $A_{\rm{B}}$, of E1 transitions with $A_{\rm{B}}$ > 10$^0$ s$^{-1}$. Right panel: Same as the left panel but for cancellation factor, CF. $LS$-allowed and $LS$-forbidden transitions are marked in blue and red, respectively.}
    \label{fig:dT}
\end{figure*}

\begin{figure}
    \centering
    \includegraphics[width=0.48\textwidth,clip]{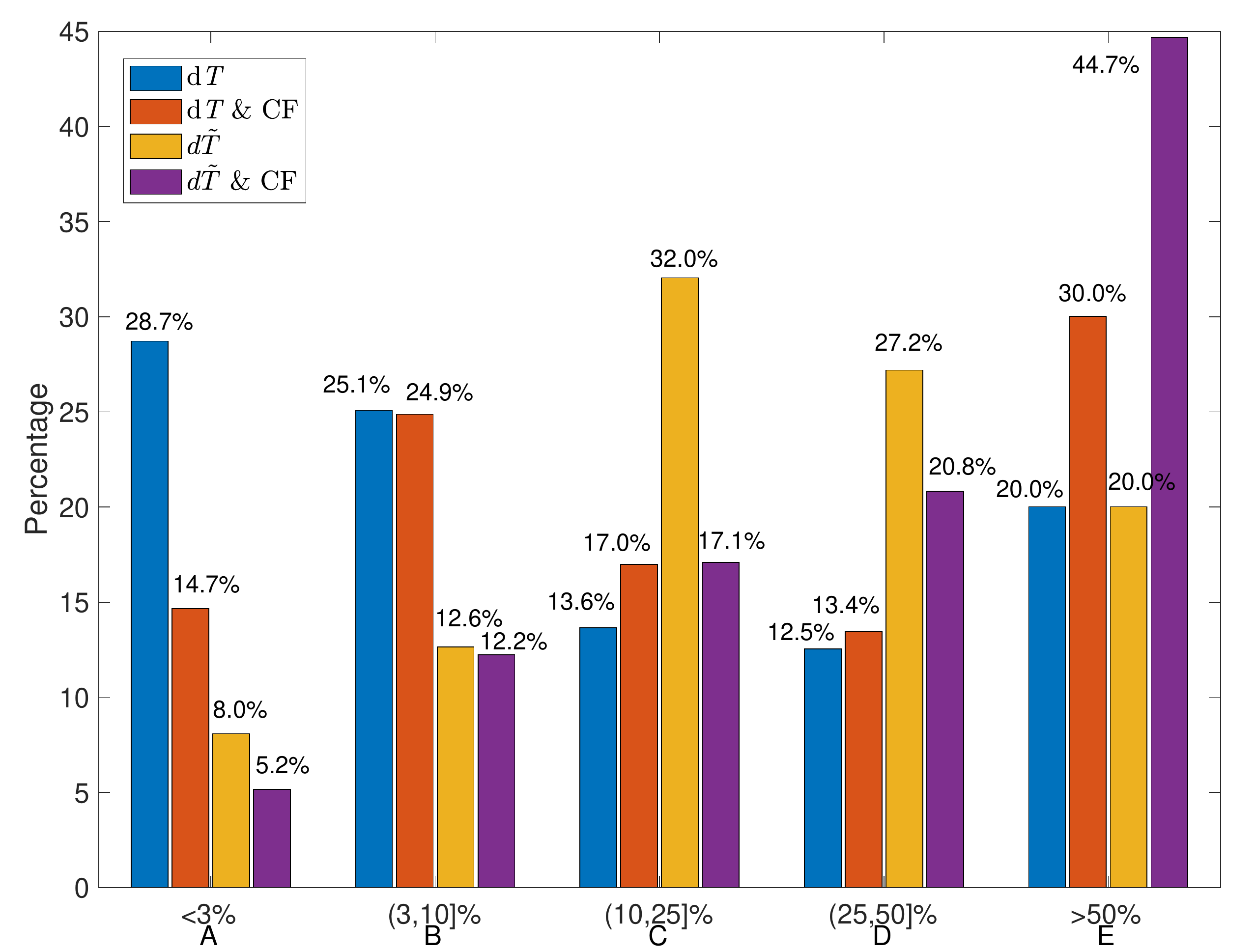}
    \caption{Percentage fractions of all transitions in \oi{} in different uncertainty categories: \re{A (Uncertainty $\le$ 3\%), B (3\% $<$ Uncertainty $\le$ 10\%), C (10\% $<$ Uncertainty $\le$ 25\%), D (25\% $<$ Uncertainty $\le$ 50\%), and E (Uncertainty $>$ 50\%),} for the uncertainty estimate based on d$T$ values only (blue), d$T$\&CF values (red), d$\tilde{T}$ (orange), and d$\tilde{T}$\&CF values (purple).}
    \label{fig:accuracy}
\end{figure}

There are a number of methods being used for estimation of uncertainties of 
calculated transition rates \citep{2014Atoms...2...86K,Fischer2009,Ekman2014,ELSAYED2021107930,Gaigalas_2020}. 
However, the estimation of uncertainties of calculated transition rates is not trivial and different methods may only applicable to 
specific systems and ionisation stages. In this work, as discussed in Sec. \ref{sec:dT&CF} (see Eqs. \eqref{eq:dT} and \eqref{eq:CF}), 
we attempted to evaluate the uncertainties using the d$T$ and CF values.

Fig. \ref{fig:dT} shows the scatterplots of d$T$ (left panel) and CF (right panel, in Babushkin gauge) versus $A$ (in Babushkin gauge).
Note that the weak transitions with transition rates $A$ < 10$^0$ s$^{-1}$ are neglected in the figure due to the fact that these weak transitions tend to be of lesser astrophysical importance, either for opacity calculations, or for spectroscopic abundance analyses.
In the figure, we depict the $LS$-allowed and $LS$-forbidden transitions in different colours.
Overall, as can be seen from Fig. \ref{fig:dT}, stronger transitions with larger $A$ rates or $LS$-allowed transitions are always associated with smaller d$T$ and larger CF values, which indicate that these transitions \re{have} small uncertainties. 
However, the weak transitions, which are mostly the $LS$-forbidden intercombination transitions, are associated with smaller CFs and \re{are} strongly affected by cancellation\re{s}.
The mean d$T$ (CF) for all E1 transitions shown in Fig. \ref{fig:dT} is \re{17.4\% (0.27)}.

To better display the d$T$ and CF parameters of the computed transitions rates and their distribution in relation to the magnitude of the transition rates $A$, we organised the transitions into four groups (g1 - g4) based on the magnitude of $A$ values, as shown in Table \ref{tab:statistic}. 
The first three groups contain the weak transitions with $A$ up to 10$^6$ s$^{-1}$, while the last group contains the strong transitions with $A \geq$ 10$^6$ s$^{-1}$.
The average value of the $\langle \rm{d}\it{T} \rangle$ is given for each group. The $\langle \rm{d}\it{T} \rangle$ is only 2.29\% for the fourth group transition, which indicates a very high accuracy achieved for the strong transitions with $A \geq$ 10$^6$ s$^{-1}$.
In addition, the statistical analysis of the proportions of transitions with CF > 0.05 and/or different d$T$ values, that is, d$T$ < 20\%, d$T$ < 10\%, and d$T$ < 5\%, for each group of transitions is also performed and shown in the last three columns of Table \ref{tab:statistic}.

Based on the d$T$ and CF values, we estimated the accuracy class for each transition using four procedures.
The first one adopts the d$T$ value defined in Eq. \eqref{eq:dT} as the uncertainty for each transition rate. 
For the second procedure, the definition of d$T$ and CF for each of the accuracy classes are presented in Table \ref{tab:accuracy}.
Furthermore, in the third approach, we organised the transitions into six groups based on the magnitude of $A$ values, that is, $A$ < 10$^{-2}$ s$^{-1}$, 
10$^{-2}$ $\le A$ < 10$^{1}$ s$^{-1}$, 10$^{1}$ $\le A$ < 10$^{3}$ s$^{-1}$, 10$^{3}$ $\le A$ < 5$\times$10$^{5}$ s$^{-1}$, 
5$\times$10$^{5}$ $\le A$ < 3.5$\times$10$^{6}$ s$^{-1}$, and $A \ge$ 3.5$\times$10$^{6}$, and calculated the averaged 
d$T_\mathrm{av}$ for each group. Then we defined d$\tilde{T}$ = max(d$T$, d$T_\mathrm{av}$) to replace the d$T$ as the uncertainty of each particular transition rate.
Finally, the fourth procedure employs the definition of d$\tilde{T}$ and CF given in Table \ref{tab:accuracy} as the uncertainty indicator.
The statistical analysis of the number of transitions belonging to \re{a} specific accuracy class is performed\re{. The} percentage fractions
obtained from the four methods are shown in Fig. \ref{fig:accuracy}.

From the comparison between the d$T$ and d$T\&$CF methods, we can see that the accuracy of some \re{of the A} class transitions is degraded according to the value of CF, 
for example, the percentage fraction is decreased from 28.\re{7}\% to 14.7\% for \re{the A} class transitions.
Compared to the other two indicators, d$\tilde{T}$ and d$\tilde{T}\&$CF predicted rather low percentage fractions of transitions in high-accuracy category 
having uncertainties less than 10\%, which are 8.0\% and 5.2\%, respectively.
From Fig. \ref{fig:accuracy}, using the d$T$ values only for accuracy estimations might underestimate the uncertainties. 
As concluded in \cite{Ekman2014}, d$T$ is a reliable indicator of uncertainties of transition rates, especially for $LS$-allowed transitions; while for $LS$-forbidden
transitions, d$T$ can be used as uncertainties indicators if averaging over a large sample.
Therefore, the d$\tilde{T}$ and d$\tilde{T}\&$CF indicators may \re{yield a} "safer" uncertainty estimation of the calculated transition rates, although they may overestimate the uncertainties, especially for the strong $LS$-allowed transitions.

The accuracy classes predicted from \re{the} d$\tilde{T}$ procedure are given in Table \ref{tab:trdata}.
A statistical analysis is performed on the distributions of accuracy classes\re{,} and the results are shown in Table \ref{tab:accuracy}.
Overall, \re{80} (205) out of 989 computed transitions in this work \re{have a} uncertainty of < 3\% (< 10\%) \re{and assigned
to accuracy class A (B).}
\re{Among} the \re{80} transitions belonging to the A accuracy class, all are rather strong, with $A \geq$ 10$^6$ s$^{-1}$ (g4). All the transitions in g4 are associated with uncertainties less than 25\%; while for \re{group} g1, most \re{transitions} are assigned to accuracy classes \re{C,} D or E.

\section{The solar oxygen abundance}

The solar oxygen abundance is still under heated debate. It has
undergone a major downward revision in the past decades, from 
$\log{\epsilon_{\mathrm{O}}}\equiv\log{N_{\mathrm{O}}/N_{\mathrm{H}}}+12$ = 8.93 in \cite{1989GeCoA..53..197A} to 8.66 in \cite{2005ASPC..336...25A}. 
Recent estimates can be separated into ``low'' values of 
$\log{\epsilon_{\mathrm{O}}}$ = 8.67-8.71
\citep{2018A&A...616A..89A,2021A&A...656A.113A,2021A&A...653A.141A}, and ``intermediate'' values of
8.73-8.77
\citep{2015A&A...579A..88C,2015A&A...583A..57S,2021MNRAS.508.2236B,2022A&A...661A.140M}, with ``high''
values in the range 8.80-8.90 sometimes also suggested 
\citep[][]{2015A&A...577A..25S,2020A&A...643A.142C}.

\input{abundancelines}

One of the differences between the recent studies of
\cite{2018A&A...616A..89A} and \cite{2021A&A...653A.141A} (low oxygen abundances)
and \cite{2021MNRAS.508.2236B} and \cite{2022A&A...661A.140M}
(intermediate oxygen abundances)
are the transition probabilities for their \oi{} lines.
In the former case, the 
transition probabilities for the \oi{} 615.8 nm, 777 nm, 844.7 nm and 926.9 nm features were taken from the NIST-ASD \re{and} are based on the atomic data of \cite{Hibbert_1991}.
On the other hand, the latter two studies, based solely on the \oi{} 777 nm triplet, draw on 
calculations from \cite{Civi__2018} computed using the quantum defect theory (QDT)\re{,} as well as from 
\cite{2022A&A...665A..18B}
based on an average over results from multiple methods.

Table~\ref{tab:abund} presents the the permitted lines that have been adopted by different
solar oxygen abundance analyses \citep{2004A&A...417..751A,2008A&A...488.1031C,2021A&A...653A.141A} . 
The log($gf$) values for these lines from various calculations are given in the table. 
It is interesting to note that, in all cases, the log($gf$) values from the NIST-ASD (based on \citealt{Hibbert_1991}) are
systematically larger than the other theoretical results. Compared to our results in the Babushkin gauge, they are between $0.01$ to $0.02\,\mathrm{dex}$ too small. Also, the QDT values of \cite{Civi__2018} are systematically smaller than the values from the other calculations; the differences with our results are between $0.01$ to $0.03\,\mathrm{dex}$.
In both cases, the differences are \re{the} largest for the \oi{} 777 nm triplet.

\begin{figure}
    \centering
    \includegraphics[width=0.48\textwidth,clip]{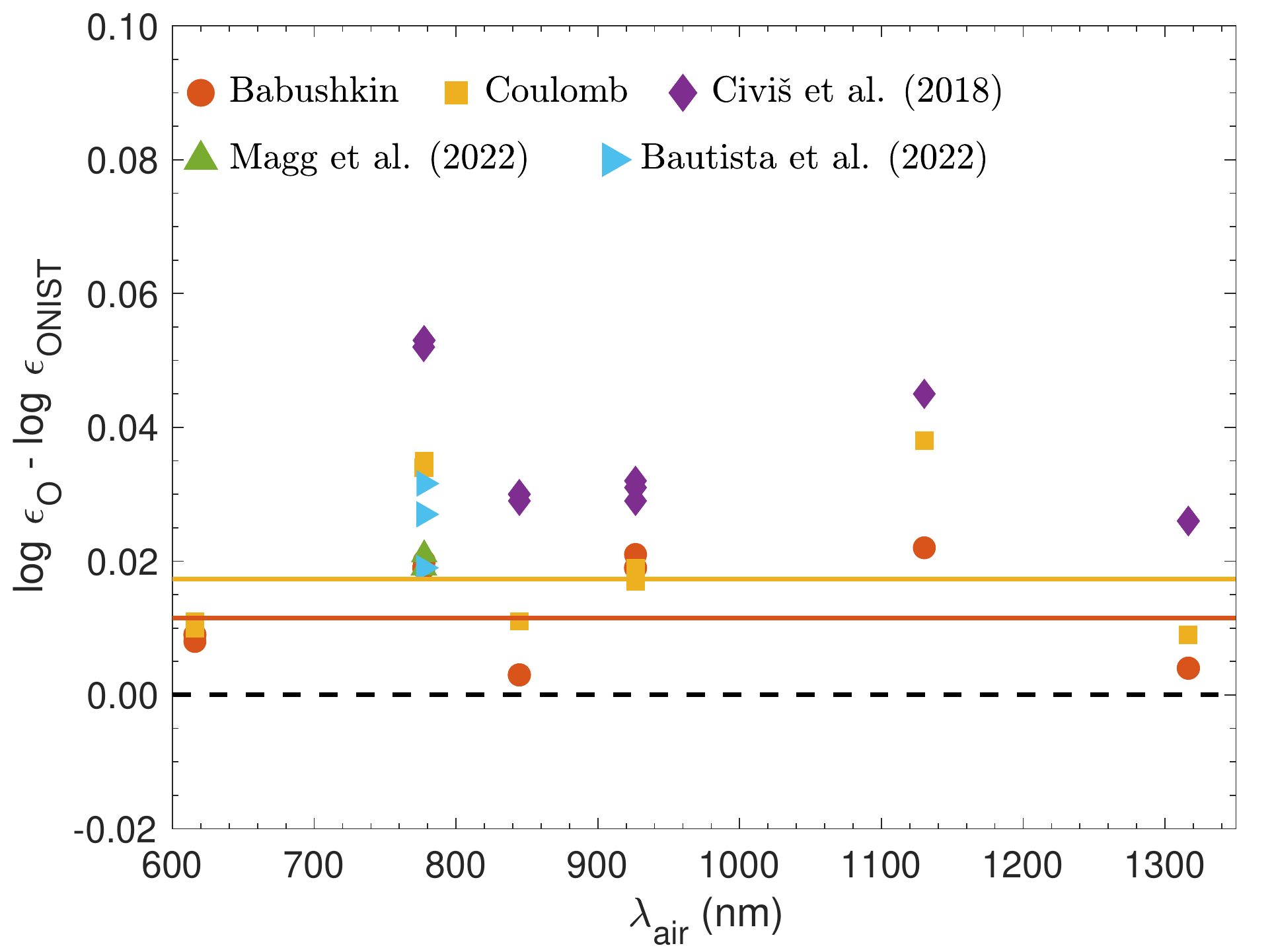}
    \caption{Solar oxygen abundance differences inferred using different theoretical transition data given in Table \ref{tab:abund}, relative to the NIST-ASD values. Red circle: Babushkin gauge from present work. Orange square: Coulomb gauge from present work. Purple diamond: \cite{Civi__2018}. Green up triangle: \cite{2022A&A...661A.140M}. Blue right triangle: \cite{2022A&A...665A..18B}.
    Red and blue solid horizontal lines illustrate the mean result for the Babushkin gauge and Coulomb gauge, respectively. Note we give equal weight\re{s} to \re{all} $LS$ feature\re{s}.
   % \amat{Unlike for C and N, here for O1 we gave almost all weight to just one line, the 777nm.  Moreover, the three components were fit simultaneously.  So, the right panel is not very informative in this case... Would it be OK to just show the left panel?  That would illustrate the main point (the effect of new loggfs on oxygen abundances) while in the text we can discuss the solar value as we have done already. You could add back again Civis and Bautista... and Magg if you like!}
   }
    \label{fig:abund}
\end{figure}

Fig.~\ref{fig:abund} illustrates the results in Table~\ref{tab:abund}
in terms of differences to the solar oxygen abundance via
\begin{equation}
\rm \Delta log~\epsilon = - \Delta~log~(\it gf) = -\left(\rm log~(\it gf)_{\rm new} - \rm log~(\it gf)_{\rm orig}\right),
    \label{eq:abund1}
\end{equation}
where in this case $\rm log~(\it gf)_{\rm orig}$ corresponds to the log($gf$) values from \cite{Hibbert_1991} via the NIST-ASD. 
One sees a systematic shift upwards, by $0.01\re{1}\,\mathrm{dex}$ \re{on} average in the Babushkin gauge, and $0.01\re{7}\,\mathrm{dex}$ \re{on} average in the Coulomb gauge. 
In particular,
for the \oi{} 777 nm triplet, \re{to} which \cite{2021A&A...653A.141A} \re{gave} the \re{largest} weight \re{among} their \re{selected} atomic lines and for which they adopt
$\log{\epsilon_{\mathrm{O}}}$= 8.69 via the analysis of \cite{2018A&A...616A..89A}, the inferred abundance increases to $\log{\epsilon_{\mathrm{O}}}$ = 8.71. This value remains in good agreement
with \re{the one} those authors obtained from forbidden lines 
(8.70) and from molecular lines (8.70; see also \citealt{2021A&A...656A.113A}), and within the quoted uncertainty of $0.04$.

The result of \cite{2018A&A...616A..89A} for the \oi{} 777 nm triplet
is $0.06-0.08\,\mathrm{dex}$ lower than that found by 
\cite{2021MNRAS.508.2236B} (8.75$\pm$0.03) and \cite{2022A&A...661A.140M} (8.77$\pm$0.04) 
from the same spectral feature.  
As illustrated in Fig.~\ref{fig:abund}, the difference in 
$\rm log~(\it gf)$ from the multi-method results of \cite{2022A&A...661A.140M} and \cite{2022A&A...665A..18B},
compared to that adopted in \cite{2018A&A...616A..89A}, amounts to around
$0.02-0.03\,\mathrm{dex}$.
These differences are significant, but \re{they are} not the dominant reason for the
$0.06-0.08\,\mathrm{dex}$ discrepancies.
Further discussion of the origins of these discrepancies may be found in Sec.~4.4 of \cite{2021A&A...656A.113A}.

Our overall advocated solar oxygen abundance is $8.70\pm0.04$, after taking into account permitted \oi{}, forbidden [\oi{}], and molecular OH lines, and adopting the same (systematic) uncertainty \re{as} given in \cite{2021A&A...653A.141A}. This falls in the ``low'' range of values as defined above, and does little to alleviate the solar modelling problem. Rather, the solution to this long-standing problem may in part derive from improvements to the theoretical modelling \citep{2015Natur.517...56B,2017A&A...607A..58B,2019A&A...621A..33B,2019ApJ...881..103Z,2019ApJ...873...18Y,2022ApJ...939...61Y}.

\section{Conclusions}
Extended atomic data including energy levels, lifetimes, and transition data of E1 transitions, are computed and provided for \oi{} using MCDHF and RCI methods. These data are indispensable for reliable solar and stellar spectroscopic analyses.

We have performed extensive comparisons of the computed transition data with other theoretical and experimental results. 
The agreement between the computed transition data and the respective results from the NIST-ASD \re{is} rather good for most of the transitions, especially for transitions with $A$ $\geq$ 10$^5$ s$^{-1}$ or log($gf$) > -1.5. 266 out of the 380 selected transitions available in the NIST-ASD are in agreement \re{with the latter} within 20\%.
\re{At the same time}, for weaker transitions with $A$ < 10$^5$ or log($gf$) < -1.5, the \re{discrepencies} between different theoretical methods \re{display a} much wider scatter. 
In particular, for transitions with 10$^3$ $\leq A \leq $ 10$^5$ s$^{-1}$ or -3.5 $\leq$ log($gf$) $\leq$ -2, the transition data in \re{the} Coulomb gauge are in better agreement with the other theoretical values than those in \re{the} Babushkin 
gauge; most of them are transitions involving high Rydberg states, for which we recommend the radiative data in \re{the} Coulomb gauge from this work.
The computed lifetimes and oscillator strengths, {log($gf$)}, have also been compared with available results from experimental measurements. Our computed values (in either gauge)
are in overall good agreement with the measured values. 

In addition, we used four methods to estimate the
uncertainties of the computed transition probabilities, 
based on the relative differences of the computed 
transition rates in the Babushkin and Coulomb gauges, which is given 
by the quantity d$T$, and CF. 
Based on the accuracy classes predicted from max(d$T$, d$T_\mathrm{av}$),
205 out of 989 computed transitions are with uncertainty less than 10\% 
and assigned to the accuracy classes \re{A or B}.
All of the computed transitions belonging to the AA accuracy class
are rather strong with $A \geq$ 10$^6$ s$^{-1}$.
All the transitions with $A \geq$ 10$^6$ s$^{-1}$ are associated with uncertainties less than 25\%; while for weak transitions with $A$ < 10$^2$ s$^{-1}$ , most are assigned to accuracy classes \re{D} or E.

Finally, the \re{impact} of the new atomic data on oxygen abundance analyses \re{was} \re{analysed} \re{by applying corrections $\rm \Delta log~\epsilon = - \Delta~log~(\it gf)$ to literature abundances.}. 
In general, the transition probabilities for typical \oi{}
lines are underestimated by \cite{Civi__2018}, and overestimated by \cite{Hibbert_1991}. For the 
\oi{} 777 nm triplet the differences with respect to our results in the Babushkin gauge are $-0.03\,\mathrm{dex}$ and $+0.02\,\mathrm{dex}$, respectively.
Our transition probabilities combined with the analysis of the \oi{} 777 nm triplet presented by \cite{2018A&A...616A..89A} suggests $\log{\epsilon_{\mathrm{o}}}=8.71$ for the Sun.
This is in excellent agreement with low-excitation forbidden lines as well as molecular lines, and, overall,
we advocate $\log{\epsilon_{\mathrm{o}}}=8.70\pm0.04$.

\section{Acknowledgements}
AMA gratefully acknowledges support from the Swedish Research Council (VR 2020-03940). MCL would like to acknowledge the support from the Guangdong Basic and Applied Basic Research Foundation
(2022A1515110043). We would like to thank the anonymous referee for his/her useful comments that helped improve the original manuscript.
\bibliographystyle{aa}
\bibliography{refs}

%%%%%%%%%%%%%%%%% APPENDICES %%%%%%%%%%%%%%%%%%%%%

\begin{appendix}
\section{Energy levels and transition data for \oi{}.}

\input{levels}

\input{tr_data}

\end{appendix}

\end{document}

%% file: scheme.tex
\begin{table*}
    \caption{Summary of the computational schemes for \oi{}.}
    \centering
    \begin{tabular}{ccccccccccccc}
    \hline\midrule
   Parity && MR in MCDHF   && MR in RCI && AS && $N_\mathrm{{CSFs}}$ \\
    \midrule
       &&   ${{2s^22p^3np~(n=2-6)}}$,     && ${2s^22p^3np~(n=2-8)}$, ${2s^22p^3nf~(n=4,5)}$,&&                              &&               \\  
even   &&    ${2s^22p^3nf~(n=4,5)}$  &&  ${2s2p^4ns~(n=3-7)}$, ${2s2p^4nd~(n=3-6)}$ &&   \{$12s, 12p, 11d, 11f, 9g, 7h$\}  &&  12 641 532   \\  
\midrule                                                                                                                    
       &&   ${{2s^22p^3ns~(n=3-6)}}$,  && ${{2s^22p^3ns~(n=3-7)}}$, &&                              &&               \\
odd    &&   ${{2s^22p^3nd~(n=3-5)}}$  && ${{2s^22p^3nd~(n=3-5)}}$ &&   \{$12s, 12p, 11d, 11f, 9g, 7h$\}  &&  7 683 274    \\  
\bottomrule
    \end{tabular}
    \label{tab:MR}
    \tablefoot{MR and AS, respectively, denote the multireference and the active sets of orbitals used in the calculations. $N_\mathrm{CSFs}$ are the numbers of generated CSFs in the final RCI calculations.}
\end{table*}

%% file: tau_exp.tex
\begin{table*}
\begin{center}
    \caption{Comparison of lifetimes in both Babushkin (B) and Coulomb (C) gauges, B/C, with other theoretical (Other Theo.) and experimental (Exp.) results.}
\label{tab:tauexp}
\begin{tabular}{llcccccccccccccccc}
\hline\midrule
& &   \multicolumn{5}{c}{lifetime} \\
\cmidrule{3-7}
State    &   Unit  &  Present    &&     Other Theo.      && Exp. \\
\midrule
${2p^33s~^5S^o}$  &  $\mu$s &   209/213     & &     528$^a$, 202/235$^d$, 200$^e$         &&      170$\pm$25$^f$, 185$\pm$10$^g$,           \\
                         &         &               & &                                  &&      180$\pm$5$^h$, 185$\pm$30$^i$          \\
${2p^33s~^3S^o}$  &  ns     &   1.70/1.69   & &     1.76$^a$,1.61$^b$, 1.73$^c$, 1.63$^e$  &&      2.4$\pm$0.3$^j$, 1.7$\pm$0.3$^k$, 1.7$\pm$0.2$^l$, 1.82$\pm$0.05$^r$,      \\
                         &         &               & &                                  &&      1.79$\pm$0.17$^p$, 1.70$\pm$0.15$^m$, 1.70$\pm$0.14$^n$, 1.8$\pm$0.27$^o$  \\
${2p^33s~^3D^o}$  &  ns     &   4.21/4.15   & &     4.19$^b$, 4.46$^c$           &&      3.94$\pm$0.22$^p$, 5.0$\pm$0.4$^q$, 4.5$\pm$0.675$^o$                           \\
${2p^33s~^1D^o}$  &  ns     &  1.86/1.85    & &     &&  1.77$\pm$0.14$^p$, 2.01$\pm$0.12$^q$ \\
${2p^34s~^3S^o}$  &  ns     &   5.74/5.67   & &     5.33$^a$, 5.24$^b$, 5.04$^c$ &&              4.0$\pm$0.6$^o$                                        \\
${2p^35s~^3S^o}$  &  ns     &   13.55/13.35 & &                                  &&      17$\pm$3$^s$, 6.0$\pm$0.9$^o$                                  \\
${2p^36s~^3S^o}$  &  ns     &   26.82/25.22 & &                                  &&      24$\pm$3$^s$                                  \\
${2p^33p~^3P}$    &  ns     &   32.12/32.73 & &     32.70$^a$, 29.68$^b$                    &&      36$\pm$4$^s$, 39.1$\pm$1.4$^t$, 40$\pm$3$^u$                           \\
${2p^34p~^3P}$    &  ns &  182.3/181.8  && 175.4$^b$  && 153$\pm$10$^u$, 161$\pm$19$^v$ \\
${2p^34p~^5P}$    &  ns &  200.2/212.0  && 189.7$^b$ && 193$\pm$10$^u$, 194$\pm$19$^v$ \\
${2p^34d~^5D^o}$  &  ns     &   72.15/73.21 & &     72.20$^b$, 96.64$^c$         &&      96$\pm$4$^u$, 95$\pm$9$^v$                                  \\
${2p^34d~^3D^o}$  &  ns     &   15.31/15.41 & &     16.85$^b$, 12.91$^c$         &&      23$\pm$3$^s$, 20$\pm$3$^o$, 80$\pm$10$^u$                                  \\
${2p^35d~^3D^o}$  &  ns     &   31.13/32.12 & &                                  &&      36$\pm$4$^s$, 30$\pm$4.5$^o$                                  \\
\hline
\end{tabular}
\\
\tablefoot{$^a$\cite{Tachiev2002A&A...385..716T};
$^b$\cite{Hibbert_1991};
$^c$\cite{Tayal_2009};
$^d$\cite{PhysRevA.102.042824};
$^e$\cite{1992A&A...265..850B};
$^f$\cite{1974PhRvA...9..568W};
$^g$\cite{1972PhRvA...5.2688J};
$^h$\cite{1978PhRvA..17.1921N};
$^i$\cite{1990ch2..book.....M};
$^j$\cite{1966ApJ...146..940S};
$^k$\cite{1968ApJ...152..695G};
$^l$\cite{1970PhLA...33..115D};
$^m$\cite{1971JOSA...61..519M};
$^n$\cite{1972CaJPh..50.2496L};
$^o$\cite{1977ApJ...214..328B};
$^p$\cite{1971ApJ...165..217S};
$^q$\cite{1974CaJPh..52.1961P};
$^r$\cite{1970PhRvA...2..397L};
$^s$\cite{1985PhRvL..55..284K};
$^t$\cite{1981CPL....82...85B,1982ApOpt..21.1419B};
$^u$\cite{1978PhyS...17..119B};
$^v$\cite{Day:81}.
Note that the experimental uncertainties for results from \cite{1977ApJ...214..328B} are given with the upper limit of the uncertainties, which is 15\%.
}
\end{center}
\end{table*}

%% file: gf_exp.tex
\begin{table*}
\begin{center}
    \caption{Comparison of $gf$ values in both Babushkin (B) and Coulomb (C) gauges, B/C, with available experimental (Exp.) results.}
\label{tab:gfexp}
\begin{tabular}{llcccccccccccccccc}
\hline\midrule
& & & & & & &  $gf$ \\
\cmidrule{4-8}
Upper        & Lower             &    $\lambda_\text{vac.}$(nm) & Present && Exp.$^{a}$ && Other Exp.  \\  
\midrule

${2p^33s~^3S^o_1}$  &   ${2p^4~^3P_0}$   &    130.603  &    0.0499/0.0501    &&        &&         0.05$\pm$0.01$^b$, 0.05$\pm$0.005$^c$, 0.047$\pm$0.0014$^d$,    \\  
                           &                           &             &                     &&                       &&       0.047$\pm$0.0047$^e$, 0.048$\pm$0.0048$^f$, 0.050$\pm$0.0050$^g$,  \\  
                           &                           &             &                     &&                       &&  0.05$\pm$0.0115$^h$, 0.052$\pm$0.0052$^i$, \\
                           &                           &             &                     &&                       &&       0.045$\pm$0.009$^j$, 0.053$\pm$0.00318$^k$  \\  
${2p^33s~^1D^o_2}$  &   ${2p^4~^1D_2}$   &    115.215  &    0.535/0.538      &&    0.49$\pm$0.039      &&        0.526$\pm$0.083$^b$, 0.526$\pm$0.055$^d$, 0.56$\pm$0.04$^e$,          \\  
                           &                           &             &                     &&                        &&       0.50$\pm$0.05$^f$, 0.51$\pm$0.026$^g$, 0.50$\pm$0.03$^l$         \\  
${2p^34s~^3S^o_1}$  &   ${2p^4~^3P_0}$   &    104.169  &    0.0082\re{7}/0.008\re{41}  &&     0.0089$\pm$0.0015  &&  \\  
${2p^34s~^3S^o_1}$  &   ${2p^4~^3P_1}$   &    104.094  &    0.024\re{9}/0.0252    &&     0.029$\pm$0.0049   &&  \\  
${2p^34s~^3S^o_1}$  &   ${2p^4~^3P_2}$   &    103.923  &    0.041\re{7}/0.0423    &&     0.048$\pm$0.0072   &&  \\  
${2p^33d~^3D^o_{1,2}}$  &   ${2p^4~^3P_1}$   &    102.743  &  0.062\re{7}/0.062\re{4}    &&     0.069$\pm$0.010    &&  \\  
${2p^33d~^3D^o_1}$  &   ${2p^4~^3P_0}$   &    102.816  &    0.020\re{9}/0.020\re{8}    &&     0.024$\pm$0.0031   &&  \\  
${2p^33s~^3D^o_1}$  &   ${2p^4~^3P_0}$   &    99.080   &    0.0570/0.057\re{7}    &&     0.052$\pm$0.0047   &&  \\  
${2p^33s~^3D^o_1}$  &   ${2p^4~^3P_1}$   &    99.013   &    0.044\re{2}/0.044\re{8}    &&     0.042$\pm$0.0042   &&  \\  
${2p^33s~^3D^o_2}$  &   ${2p^4~^3P_1}$   &    99.020   &    0.128/0.130      &&   0.11$\pm$0.0121      &&  \\  
${2p^33s~^3D^o_2}$  &   ${2p^4~^3P_2}$   &    98.865   &    0.0457/0.0463    &&     0.051$\pm$0.0071   &&  \\  
${2p^33s~^3D^o_3}$  &   ${2p^4~^3P_2}$   &    98.877   &    0.243/0.24\re{6}      &&   0.22$\pm$0.044       &&  \\  
\hline
\end{tabular}
\tablefoot{$^a$\cite{Goldbach1994A&A...284..307G};
$^b$\cite{1968ApJ...152..695G}; 
$^c$\cite{1970PhLA...33..115D}; 
$^d$\cite{1970PhRvA...2..397L}; 
$^e$\cite{1971ApJ...165..217S}; 
$^f$\cite{1971JOSA...61..519M}; 
$^g$\cite{1972CaJPh..50.2496L}; 
$^h$\cite{1971PhRvA...4..245O}; 
$^i$\cite{1971ApOpt..10.1288K}; 
$^j$\cite{1976Clyne}; 
$^k$\cite{1985JQSRT..34...55J}; 
$^l$\cite{1974CaJPh..52.1961P}.
The results from present calculations are adjusted to experimental wavelengths.}
\end{center}
\end{table*}

%% file: statistic.tex
\begin{table*}
    \caption{Distribution of d$T$ (in \%) and CF of the computed transition rates in \oi~depending on the magnitude of the transition rates. \re{The analysis is done based on the data adjusted to experimental wavelengths.}}
    \centering
    \begin{tabular}{clcccccccc}
    \hline\midrule
Group  & $A$  &  No. &  $\langle \rm{d}\it{T} \rangle$ &  $ \rm{d}\it{T}$ < 20\% (\& CF > 0.05)  &  $\rm{d}\it{T}$ < 10\% (\& CF > 0.05) &  $\rm{d}\it{T}$ < 5\% (\& CF > 0.05)   \\
  & (s$^{-1}$)  &   &   (\%) &  (\%)  &  (\%) &  (\%)   \\
\midrule
g1 & 10$^0$--10$^2$  &  226 &  28.32         &      58.8 (0.4)    &   38.1 (0.4)      &    23.5 (0.0)    \\              
g2 & 10$^2$--10$^4$  &  147 &  17.03         &      74.8 (24.5)   &   72.1 (23.1)     &    57.1 (12.2)   \\
g3 & 10$^4$--10$^6$  &  221 &  17.99         &      75.6 (65.2)   &   63.3 (54.3)     &    52.5 (45.7)   \\
g4 & 10$^6$--        &  166 &  2.29          &      100.0 (95.8) &   98.8 (94.6)     &    86.7 (82.5)   \\
\bottomrule
    \end{tabular}
    \label{tab:statistic}
\end{table*}

%% file: accuracy.tex
\begin{table*}
    \caption{The connection of the limits of d$T$ or d$\tilde{T}$ and CF for the accuracy classes (Acc.).}
    \centering
    \begin{tabular}{llllrrrrr}
    \hline\midrule
Acc.   &  Unc.  &  d$T$ or d$\tilde{T}$  &  CF  & $N$ & \multicolumn{4}{c}{$N_{\mathrm{acc.}}$}\\
\cmidrule{6-9}
& (\%) & (\%) & & & g1 & g2 & g3 & g4 \\
\midrule
A & $\leq$ 3 & $\leq$ 3   &  $\geq$ 0.1 & 80                                                &  0                    &   0                    &  0                   &  80  \\  
\multirow{2}{*}{B} & \multirow{2}{*}{$\leq$ 10} & > 3 \& $\leq$ 10   &  $\geq$ 0.1 & \multirow{2}{*}{125}   &  \multirow{2}{*}{0} &   \multirow{2}{*}{0}   &  \multirow{2}{*}{41}  & \multirow{2}{*}{84}  \\  
 & &  $\leq$ 3   &  < 0.1 & &\\                                                            
\multirow{2}{*}{C}& \multirow{2}{*}{$\leq$ 25} & > 10 \& $\leq$ 25   &  $\geq$ 0.1 & \multirow{2}{*}{317}  &  \multirow{2}{*}{70} &   \multirow{2}{*}{111}   &  \multirow{2}{*}{134}  & \multirow{2}{*}{2}   \\  
 & & >3 \& $\leq$ 10   &  < 0.1 & & \\                                                    
\multirow{2}{*}{D} & \multirow{2}{*}{$\leq$ 50} & > 25 \& $\leq$ 50   &  $\geq$ 0.1 & \multirow{2}{*}{269}  &  \multirow{2}{*}{110} &   \multirow{2}{*}{20}  &    \multirow{2}{*}{19} &  \multirow{2}{*}{0} \\
 & & or >10 \& $\leq$ 25   &  < 0.1 & & \\                                                 
\multirow{2}{*}{E} & \multirow{2}{*}{$>$ 50} & > 50   &   & \multirow{2}{*}{198}                             &  \multirow{2}{*}{46} &   \multirow{2}{*}{16}  &   \multirow{2}{*}{27} &  \multirow{2}{*}{0} \\
 & & > 25 \& $\leq$ 50   &  < 0.1 & & \\
\bottomrule
    \end{tabular}
    \label{tab:accuracy}
    \tablefoot{$N$ is the total number of computed transitions belonging to \re{a} specific accuracy class obtained from \re{the} d$\tilde{T}$ indicator. $N_{\mathrm{acc.}}$ is the number of transitions belonging to \re{a} specific accuracy class in each transition group defined in Table \ref{tab:statistic}. Note that the CF in \re{the} Babushkin gauge is used in this statiscal analysis. The A, B, C, D, and E classes includes, respectively, the \{A, A+, and AA\}, \{B+ and B\}, \{C+ and C\}, \{D+ and D\}, and E classes as defined by the NIST-ASD. The corresponding uncertainty (Unc.) limits are shown in the second column.}
\end{table*}

%% file: abundancelines.tex
\begin{table*}
\begin{center}
    \caption{{Allowed} \oi{} lines used as oxygen abundance diagnostics.}
\label{tab:abund}
\begin{tabular}{llcccccccccccc}
\hline\midrule
& & & \multicolumn{6}{c}{$\log gf$} \\
\cmidrule{4-10}
Upper        & Lower             &    $\lambda_\text{air}$(nm) & \multicolumn{2}{c}{Present} & NIST$^{(a)}$ & QDT$^{(b)}$ & Multi-method$^{(c)}$  & Multi-method$^{(d)}$\\
\cmidrule{4-5}
  &   &  &   B & C  & & & & \\
\midrule
${2p^34d~^5D_2^o}$  &  ${2p^33p~^5P_3}$  &  615.815 & -1.849  &  -1.851   & -1.841    &  && \\
${2p^34d~^5D_3^o}$  &  ${2p^33p~^5P_3}$  &  615.817 & -1.004  &  -1.006   & -0.995   &  &&\\
${2p^34d~^5D_4^o}$  &  ${2p^33p~^5P_3}$  &  615.819 & -0.418  &  -0.420   & -0.409   &  &&\\
${2p^33p~^5P_3}$  &  ${2p^33s~^5S_2^o}$  &  777.194 &  0.350  &   0.335   &  0.369    & 0.317   & 0.350   & 0.350$\pm$0.021   \\  
${2p^33p~^5P_2}$  &  ${2p^33s~^5S_2^o}$  &  777.417 &  0.204  &   0.189   &  0.223    & 0.170   & 0.204   & 0.196$\pm$0.022   \\  
${2p^33p~^5P_1}$  &  ${2p^33s~^5S_2^o}$  &  777.539 & -0.018  &  -0.033   &  0.002    &-0.051   & -0.019   & -0.0296$\pm$0.021  \\  
${2p^33p~^3P_0}$  &  ${2p^33s~^3S_1^o}$  &  844.625 & -0.466  &   -0.474    &   -0.463    & -0.493 & && \\
${2p^33p~^3P_2}$  &  ${2p^33s~^3S_1^o}$  &  844.636 & 0.233  &   0.225    &    0.236   & 0.206 & && \\
${2p^33p~^3P_1}$  &  ${2p^33s~^3S_1^o}$  &  844.676 &  0.011  &  0.003     &  0.014     &-0.015& && \\
${2p^33d~^5D_2^o}$  &  ${2p^33p~^5P_3}$  &  926.583 & -0.739  &  -0.737     & -0.718     &-0.750& && \\
${2p^33d~^5D_3^o}$  &  ${2p^33p~^5P_3}$  &  926.594 &  0.106  &   0.108     &  0.125    & 0.096& && \\
${2p^33d~^5D_4^o}$  &  ${2p^33p~^5P_3}$  &  926.601 &  0.693  &   0.694     &  0.712    & 0.681& && \\
${2p^34s~^5S_2^o}$  &  ${2p^33p~^5P_3}$  & 1130.238  &  0.056  &  0.040      &   0.078   & 0.033 & && \\
${2p^34s~^3S_1^o}$  &  ${2p^33p~^3P_1}$  &  1316.389 &   -0.258 &   -0.263     &  -0.254    & -0.280 & && \\
${2p^34s~^3S_1^o}$  &  ${2p^33p~^3P_2}$  & 1316.485  &  -0.036 &   -0.041     &  -0.032     & -0.058& && \\
${2p^34s~^3S_1^o}$  &  ${2p^33p~^3P_0}$  & 1316.511  &  -0.735 & -0.740       &  -0.731    & -0.757 & && \\

\hline
\end{tabular}\\
\tablefoot{$^{(a)}$\cite{NIST_ASD};
$^{(b)}$\cite{Civi__2018};
$^{(c)}$\cite{2022A&A...661A.140M};
$^{(d)}$\cite{2022A&A...665A..18B}. Shown are the upper and lower configurations, 
\re{experimental} wavelength (nm~in air), and oscillator strengths
obtained from various calculations. Note {that} the log($gf$) values in the penultimate column for the 777-triplet lines given by \cite{2022A&A...661A.140M} are based
on the results calculated with both the GRASP
and AUTOSTRUCTURE code. The values in the last column {reported} by \cite{2022A&A...665A..18B} are obtained by averaging
over results from multiple methods (MCDHF, R-matrix, pseudo-relativistic Hartree-Fock method + core-polarization effects, and CI).The log($gf$) results from {the} present calculation are adjusted using experimental wavelengths. 
}
\end{center}
\end{table*}

%% file: levels.tex
\begin{table*}
\caption{\label{tab:energy} Energy levels (in cm$^{-1}$) and lifetimes (in s; given in Babushkin ($\tau_B$) and Coulomb ($\tau_C$) gauges) for \oi.} 
\begin{tabular}{llllllllllllllllll}
\hline\midrule
No. & State   && $E_\mathrm{MCDHF} $ && $E_\mathrm{NIST} $ && ${\Delta E}$ && $\tau_B$ && $\tau_C$  \\ 
\midrule
  1 &    ${2s^22p^4~^3P_2}$          &&      0.00  &&      0.000 &&    0     &&             &&  \\ 	
  2 &    ${2s^22p^4~^3P_1}$          &&    155.73  &&    158.265 &&  2.54    &&             &&     \\   	
  3 &    ${2s^22p^4~^3P_0}$          &&    223.26  &&    226.977 &&  3.72    &&             &&     \\   	
  4 &    ${2s^22p^4~^1D_2}$          &&  15966.11  &&  15867.862 &&  -98.25  &&             &&      \\  	
  5 &    ${2s^22p^4~^1S_0}$          &&  34053.62  &&  33792.583 &&  -261.04 &&             &&       \\ 	
  6 &    ${2s^22p^33s~^5S_{2}^o}$    &&  73705.01  &&  73768.200 &&  63.19   &&  2.0918E-04 &&   2.1296E-04  \\	  
  7 &    ${2s^22p^33s~^3S_{1}^o}$    &&  76728.03  &&  76794.978 &&  66.95   &&  1.7020E-09 &&   1.6911E-09  \\	  
  8 &    ${2s^22p^33p~^5P_{1}}$      &&  86455.25  &&  86625.757 &&  170.51  &&  2.9107E-08 &&   3.0123E-08  \\	  
  9 &    ${2s^22p^33p~^5P_{2}}$      &&  86457.22  &&  86627.778 &&  170.56  &&  2.9094E-08 &&   3.0109E-08  \\	  
 10 &    ${2s^22p^33p~^5P_{3}}$      &&  86460.80  &&  86631.454 &&  170.65  &&  2.9070E-08 &&   3.0085E-08  \\
 -- & -- && --&& -- && --&& --&& --\\
\bottomrule
\end{tabular}
\tablefoot{Energy levels are given relative to the ground state and compared with NIST-ASD data \citep{NIST_ASD}. The differences ${\Delta E}$ between $E_\mathrm{MCDHF}$ and $E_\mathrm{NIST}$ values are shown in the fifth column. The full table is available online.}
%\end{longtable}
\end{table*}

%% file: tr_data.tex
\begin{table*}
\tiny
\caption{\label{tab:trdata} Electric dipole transition data for \oi{}.}
\centering
\begin{tabular}{llccccccccccccc}
\hline\midrule
Upper & Lower  & $\lambda$ (nm) & \multicolumn{2}{c}{$S$~(a.u. of a$_0^2$e$^2$)}& \multicolumn{2}{c}{$\mathrm{log}~gf$} &  \multicolumn{2}{c}{$A~(\mathrm{s}^{-1})$}  & d$T$ & \multicolumn{2}{c}{CF} \\
\cmidrule{4-5}\cmidrule{6-7}\cmidrule{8-9}\cmidrule{11-12}
 &     &  & B & C & B  & C & B & C & & B & C & Acc.\\
\midrule
  ${2s^22p^35d~^3D_{1}^o}$    &    ${2s^22p^4~^3P_{2}}$      &         94.8686  &   1.01E-03  &     9.82E-04  &    -3.489  &  -3.502    &    8.01E+05  &     7.77E+05 &    0.030  &   1.90E-03  &   3.71E-02 &  B\\
  ${2s^22p^35d~^3D_{2}^o}$    &    ${2s^22p^4~^3P_{2}}$      &         94.8686  &   1.52E-02  &     1.47E-02  &    -2.313  &  -2.327    &    7.20E+06  &     6.98E+06 &    0.030  &   3.84E-03  &   1.40E-01 &  B\\
  ${2s^22p^35d~^3D_{3}^o}$    &    ${2s^22p^4~^3P_{2}}$      &         94.8686  &   8.48E-02  &     8.23E-02  &    -1.566  &  -1.579    &    2.87E+07  &     2.79E+07 &    0.030  &   8.04E-03  &   3.74E-01 &  A\\
  ${2s^22p^35d~^5D_{3}^o}$    &    ${2s^22p^4~^3P_{2}}$      &         94.8898  &   1.37E-06  &     1.38E-06  &    -6.356  &  -6.354    &    4.66E+02  &     4.68E+02 &    0.006  &   1.32E-07  &   6.21E-06 &  C\\
  ${2s^22p^35d~^5D_{2}^o}$    &    ${2s^22p^4~^3P_{2}}$      &         94.8898  &   1.68E-08  &     1.85E-08  &    -8.269  &  -8.226    &    7.97E+00  &     8.79E+00 &    0.094  &   3.63E-09  &   1.63E-07 &  D\\
  ${2s^22p^35d~^5D_{1}^o}$    &    ${2s^22p^4~^3P_{2}}$      &         94.8898  &   3.31E-08  &     3.16E-08  &    -7.975  &  -7.994    &    2.61E+01  &     2.50E+01 &    0.044  &   2.64E-08  &   8.23E-07 &  C\\
  ${2s^22p^35d~^3D_{1}^o}$    &    ${2s^22p^4~^3P_{1}}$      &         95.0112  &   1.51E-02  &     1.47E-02  &    -2.315  &  -2.328    &    1.19E+07  &     1.16E+07 &    0.030  &   6.02E-03  &   2.46E-01 &  B\\
  ${2s^22p^35d~^3D_{2}^o}$    &    ${2s^22p^4~^3P_{1}}$      &         95.0112  &   4.54E-02  &     4.40E-02  &    -1.838  &  -1.851    &    2.14E+07  &     2.08E+07 &    0.030  &   7.97E-03  &   3.78E-01 &  B\\
  ${2s^22p^35d~^5D_{2}^o}$    &    ${2s^22p^4~^3P_{1}}$      &         95.0325  &   8.17E-07  &     8.20E-07  &    -6.583  &  -6.581    &    3.86E+02  &     3.87E+02 &    0.003  &   1.68E-07  &   7.74E-06 &  C\\
  ${2s^22p^35d~^5D_{1}^o}$    &    ${2s^22p^4~^3P_{1}}$      &         95.0325  &   8.94E-09  &     9.69E-09  &    -8.544  &  -8.509    &    7.03E+00  &     7.63E+00 &    0.078  &   4.54E-09  &   2.03E-07 &  D\\

-- & --& -- & -- & -- & --& -- & --& --& --& --& --\\
\bottomrule
\end{tabular}
\tablefoot{Upper and lower states, wavelength \re{in vacuum}, $\lambda$, line strength, $S$, weighted oscillator strength, log $gf$, transition probability, $A$, together with the relative difference between two gauges of $A$ values, $dT$, and cancellation factor, CF, are shown in the table.
Note that the wavelengths and transition parameters are adjusted to the NIST-ASD Ritz wavelength values \citep{NIST_ASD}. \re{The limits of the accuracy classes, Acc., are defined as: A $\le$ 3\%, B $\le$ 10\%, C $\le$ 25\%, D $\le$ 50\%, and E $>$ 50\%.} Only the first ten rows are shown; the full table is available online.}
\end{table*}